\documentclass[12pt]{article}
\usepackage{graphicx}
\usepackage{amsmath}
\usepackage{amssymb}
\usepackage{authblk}
\usepackage{geometry}
\usepackage[square,numbers,sort&compress]{natbib}
\usepackage[thinc]{esdiff}

\title{Dynamic roughening of cities driven by multiplicative noise}
% ...
\author[1,*]{Martin Hendrick}
\author[1,*]{Gabriele Manoli}

\affil[1]{Laboratory of Urban and Environmental Systems, \'{E}cole Polytechnique F\'{e}d\'{e}rale de Lausanne, Lausanne, Switzerland}

\affil[*]{\small Corresponding authors: gabriele.manoli@epfl.ch and martin.hendrick@epfl.ch}

\date{\today}
\begin{document}

\maketitle

\begin{abstract}
The evolution of urban landscapes is rapidly altering the surface of our planet. Yet, our understanding of the urbanisation phenomenon remains far from complete. A fundamental challenge is to describe spatiotemporal changes in the built environment. A dynamic theory of urban evolution should account for both vertical and horizontal city expansion, analogous to the dynamical behaviour of surface growth in physical and biological systems. Here we show that building-height dynamics in cities around the world are well described by a zero-dimensional geometric Brownian motion (GBM), where multiplicative noise drives stochastic fluctuations around a deterministic drift associated with economic growth. To account for intra-city correlations, we extend the GBM with spatial coupling, revealing how local interactions effectively mitigate noise-driven fluctuations and shape urban morphology. The continuum limit of this spatial model can be recasted into the Kardar–Parisi–Zhang (KPZ) equation and we find that empirical estimates of the roughness exponent are in the range of the KPZ prediction for most cities. Together, these results show that multiplicative noise, moderated by local interactions, governs the evolution of urban roughness, anchoring spatiotemporal city dynamics in a well‑established statistical‑physics framework.
\end{abstract}


\section*{Introduction}

Cities worldwide are experiencing rapid and heterogeneous growth, driven by economic expansion, demographic change, and technological development. High-resolution global datasets show that between 1985 and 2015, urban land cover increased by nearly 10{,}000~km$^2$ per year, with the fastest expansion occurring in China, India, and Africa \cite{liu2020high}. This large-scale urbanization of the Earth surface puts increasing pressure on natural ecosystems and creates sustainability challenges both locally and 
globally \cite{grimm2008global}. Hence, despite progress in our understanding of urban development at the macroscopic level \cite[e.g.][]{bettencourt2007growth,verbavatz2020growth}, there is an urgent need for quantitative models that capture the detailed spatiotemporal dynamics of urban structure in order to anticipate change, guide urban planning, and inform decision-makers with robust scientific information \cite[e.g.][]{barthelemy2019statistical}. 

Empirical research has uncovered striking regularities in urban systems, such as Zipf’s law for city size distributions \cite{gabaix1999zipf,rozenfeld2011area} and scaling laws linking urban indicators to population size \cite{bettencourt2007growth}. Classical urban economics models, including the Alonso--Muth--Mills model \cite{alonso1964location} and central place theory \cite{fischer2011central}, provide equilibrium descriptions of urban spatial structure, while economic geography frameworks incorporate increasing returns and trade costs \cite{krugman1996self}. However, these approaches often treat cities as point-like objects or assume static monocentric structures, limiting their ability to describe the actual spatiotemporal evolution of cities.

In parallel, statistical physics and complexity science have offered dynamic and spatially explicit models. Fractal analyses and diffusion-limited aggregation models reproduce features of urban morphology \cite{batty1986fractal,frankhauser1994fractalite,batty1991cities}, while correlated percolation models describe urban sprawl and merging \cite{makse1998modeling}. Reaction–diffusion frameworks have also been used to describe the horizontal expansion of urban areas \cite{raimbault2018calibration,tirico2018morphogenesis}, including the co-evolution of population and transport networks over multi-decadal timescales \cite{capel2024angiogenic}. Recent work has linked urban growth fronts to kinetic roughening: while some cities fall into Kardar–Parisi–Zhang (KPZ) or Edwards–Wilkinson universality classes \cite{najem2020kinetic}, others seem to be inconsistent with KPZ and are better captured by alternative dynamical scaling ansatz \cite{marquis2025universal}. Stochastic models of city size evolution \cite{verbavatz2020growth,reia2022spatial}, capture deviations from Zipf's law due to random shocks and finite-time effects, and cross-diffusion models have addressed interactions of density and housing prices at the intra-city level \cite{jin2023detecting}. 

Many of these models are based on multiplicative noise, which has a long history in urban and economic growth — from Gibrat’s law \cite{gibrat1931inegalites} to random-growth explanations of Zipf’s law for city sizes, economies \cite[e.g.][]{eeckhout2004gibrat,verbavatz2020growth,reia2022spatial,bettencourt2020urban} and geomorphology \cite{bonetti2017dynamic}. More generally, multiplicative noise encapsulates a wide variety of phase–transition phenomena, from depinning of interfaces to wetting and synchronization, and generates a rich zoo of universality classes \cite{munoz2003multiplicativenoisenonequilibriumphase}. In socio‑economic contexts, the same mechanism governs how resources concentrate or disperse \cite{bouchaud2000wealth}. A recent mean‑field analysis of heterogeneous random growth shows that multiplicative fluctuations alone lead to extreme localization unless sufficiently strong migration (i.e., redistribution) counteracts this tendency \cite{bernard2025mean}.

Although some studies in the economics literature addresses cities’ vertical expansion \cite[e.g.][]{barr2021growing}, most of the aforementioned physics-based research on urban growth focuses on horizontal sprawl only. Yet, recent observations indicate a global shift towards building upward \cite{frolking2024global}. Here we address this knowledge gap by introducing a physics-based framework — based on a minimalist stochastic model with multiplicative noise — that jointly reproduces vertical and horizontal growth in world cities and quantifies intra-urban volumetric dynamics. We start from a zero-dimensional Geometric Brownian Motion (GBM), which encapsulates local stochastic proportional growth in an analytically tractable way. We then introduce spatial coupling by adding an interaction term between neighbouring sites, creating a spatially extended system. Using a mean-field approximation, we study the emergence of distinct growth regimes characterised by the mean and variance of building heights. Finally, by taking the continuum limit, we arrive at a stochastic partial differential equation connected to the KPZ universality class \cite{kardar1986dynamic} via the Hopf--Cole transformation. The deterministic growth rate of our model is found to be directly linked to the local Gross Domestic Product (GDP), connecting economic productivity to the dynamics of urban structure. Our theoretical framework thus directly couples economic productivity to the spatial growth of cities, demonstrating its connection to the KPZ equation and the theory of growing interfaces in the physical and biological domains, from particle deposition to bacterial growth \cite{wakita1997self}.

\section*{Results}

As a first step, we analysed urban vertical growth through the lens of geometric Brownian motion (Eq.~\ref{eq:GBM}; see Methods), a canonical model for multiplicative processes in which increments are proportional to the current state.  Its logarithmic growth rate, defined for each city as $\gamma=\mu-\sigma^{2}/2$, collapses the two GBM parameters (deterministic drift $\mu$ and noise amplitude $\sigma$) into a single measure that is readily comparable across cities, countries, and time (Eq.~\ref{eq:gamma}; see Methods).

\begin{figure}[h]
    \centering
    \includegraphics[width=1\linewidth]{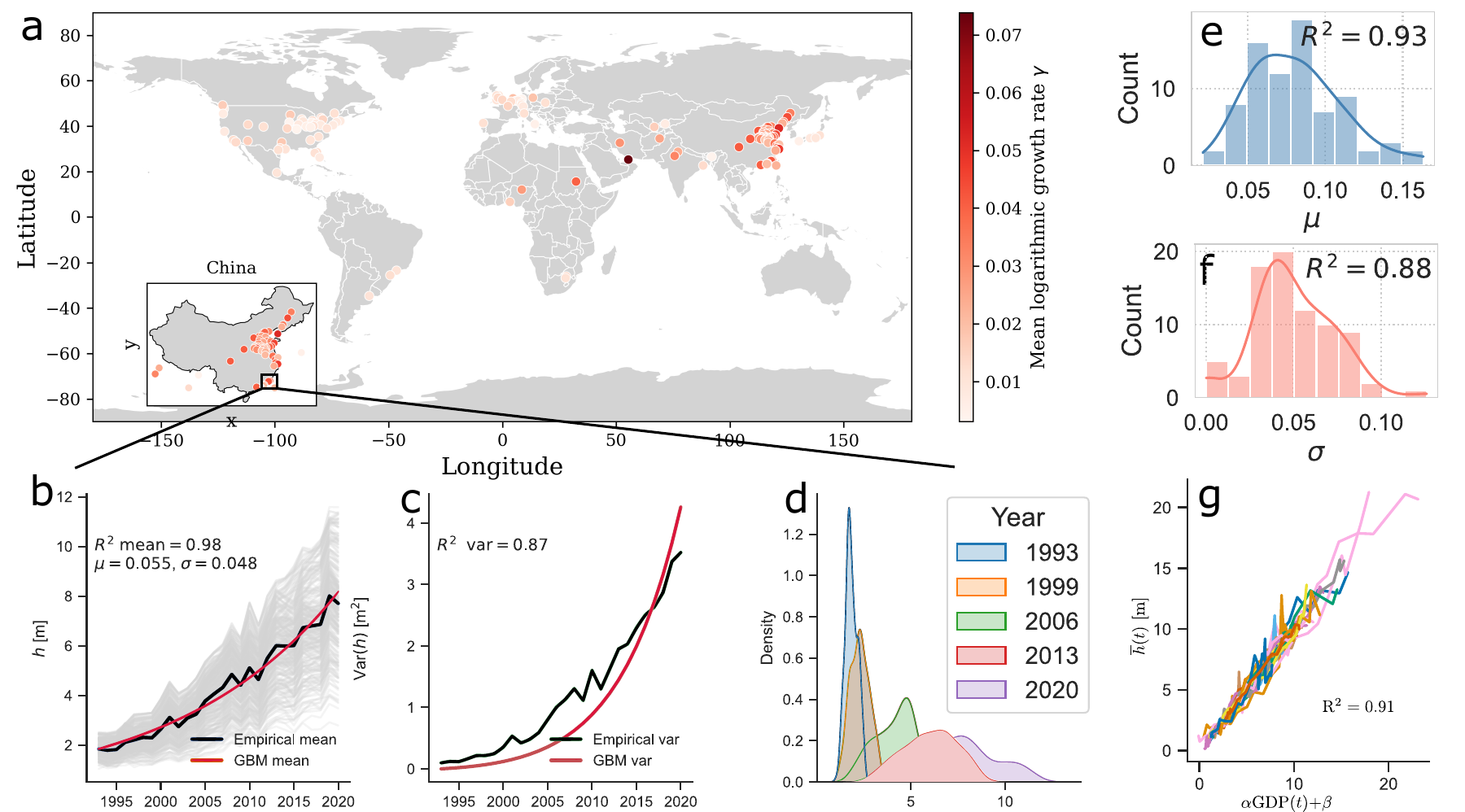}
 \caption{\textbf{GBM growth of world cities.} (a) World map of analysed cities colored by the mean logarithmic building-height growth rate $\gamma$; inset highlights East Asia. Guangzhou, China, shown as an illustrative case (highlighted by the black box in the inset). (b–d) GBM–data comparison for Guangzhou: (b) per-pixel building heights (grey) with empirical and GBM mean height overlaid (Eq.~\ref{eq:GBM_interactions_mean}); (c) variance of heights, empirical vs GBM (Eq.~\ref{eq:GBM_variance}); (d) kernel-density estimates of the height distribution for five benchmark years. (e,f) GBM parameter estimates for all East Asian cities analyzed where $R^{2}$ refers to empirical mean (f) and variance (g) compared to GBM. (g) Temporal lienar relation between city mean height $\overline{h}$ and city GDP ($\alpha,\beta$ constant determined for each city; one line represent one city.)}

    \label{fig:map}
\end{figure}

Applying the formalism to building height ($h$) observations from a sample of 500 metropolitan areas over the period 1993–2020, we find that high logarithmic growth rates cluster in China and South‑East Asia, mirroring the region’s rapid economic ascent (Fig.~\ref{fig:map}a).  In contrast, for example, U.S.\ and European cities exhibit markedly lower $\gamma$ values. In addition, we find that the logarithmic growth rate is independent of city size, implying a size-independent growth law consistent with Gibrat-type regularity (see SI Fig.~\ref{fig:gamma_size}).

Focusing on individual cities helps clarify the underlying mechanisms. Guangzhou exemplifies a fast-growing Chinese center: its empirical building height trajectory closely matches GBM predictions, where the height at each location follows a GBM path with constant parameters for the city. This holds for both the mean $\overline{h}(t)$ and the variance $\text{Var}(h) $ ($R^{2} \approx0.9$; Fig.\ref{fig:map}bc). We extended this analysis to the 116 Chinese cities and surrounding areas (Fig.\ref{fig:map}ef), yielding a pair of parameters $(\mu,\sigma)$ for each city and consistently good $R^{2}$ values for both the mean and variance fits. Remarkably, we observed a linear relationship over time between mean building height and city GDP, indicating that the intrinsic city growth rate $\mu$ is directly tied to GDP growth (Fig.~\ref{fig:map}g).

By contrast, New York City (Supplementary in Fig.~\ref{fig:NY}; or Phoenix for a less extreme example provided in Supplementary in Fig.~\ref{fig:phoenix}), exemplifies a mature USA city that deviates from GBM predictions. In particular, the variance deviates because substantial height dispersion was already present in 1993 and subsequent growth approaches a saturation plateau where fluctuations dominate (Supplementary Fig.~\ref{fig:usa_gbm_city}; a behaviour that can be reproduced by extending the GBM model to the stochastic Fisher-Komolgorov equation, see Eq.~\ref{eq:sFK} and Methods). Nevertheless, GDP continues to track the mean height reliably for most of the USA cities (Supplementary Fig.~\ref{fig:gdp_vs_hieght_usa}), underscoring that economic output reflects average vertical expansion even in late‑stage urban growth dynamics.

\begin{figure}[h]
    \centering
    \includegraphics[width=1\linewidth]{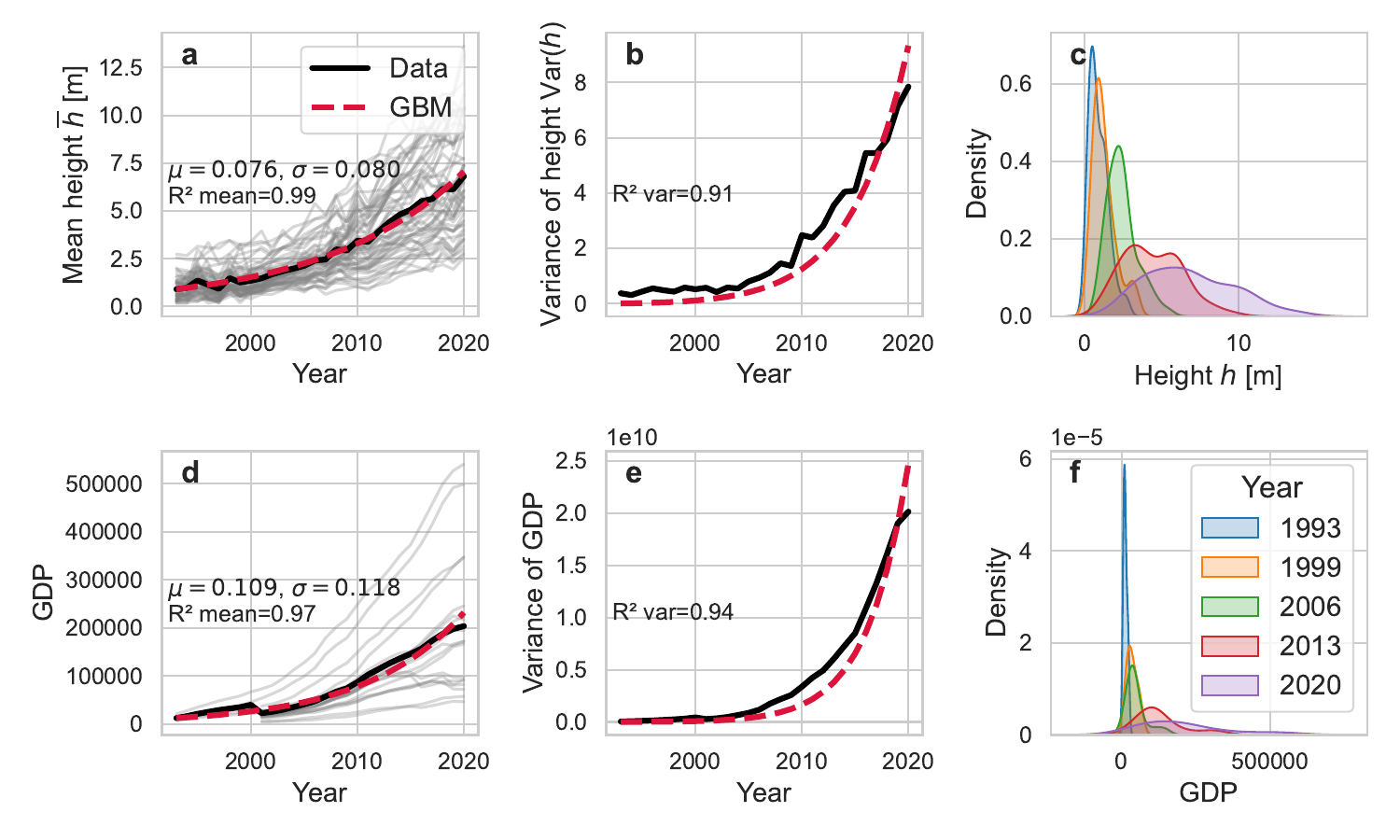}
\caption{\textbf{City‑level GBM behaviour for China.}   
\textbf{(a)} Time series of the mean building height $\overline{h}(t)$ for each city (grey) and the national average (black), compared with the GBM fit (red).  
\textbf{(b)} Corresponding height variance $\mathrm{Var}\!\bigl(h(t)\bigr)$ and its GBM prediction.  
\textbf{(c)} Kernel‑density estimates of the height distribution for five benchmark years (1993–2020).  
\textbf{(d)} National‑average GDP (black) with GBM fit (red), GDP for each city (grey; city level GDP data are only available for the main Chinese cities \cite{NBS2025}).  
\textbf{(e)} Variance of GDP and GBM prediction.  
\textbf{(f)} KDEs of the GDP distribution for the same benchmark years.  
Inset boxes list the fitted GBM parameters $(\mu,\sigma)$ and coefficients of determination $R^{2}$ for mean and variance.  The analysis extends earlier intra‑urban study to the inter‑city scale.}
\label{fig:GBM_country_level}
\end{figure}

This intra-urban toy model is also generalisable to the national scale. For China, a single drift–volatility pair reproduces the temporal evolution of the country-average building height (Fig.~\ref{fig:GBM_country_level}a–c). The same approach captures the temporal dynamics of city GDP (Fig.~\ref{fig:GBM_country_level}d–f), with drift and volatility comparable to those inferred for country-average and intra-urban building-height dynamics (Fig.~\ref{fig:map}, see also Supplementary Fig.~\ref{fig:usa_gbm_city} for the United States as a contrasting case). Taken together, these results support the self-similarity of urban growth across scales, as previously noted by \citet{bettencourt2020urban}.

Despite the effectiveness of GBM at reproducing the city- and national-scale building height dynamics observed in fast-growing cities (large $\gamma$), it does not account for the spatial correlations that exist within individual cities. To overcome this limitation, we extend the GBM model by adding local interactions between neighbouring locations resulting in a spatio-temporal stochastic growth model (see also Methods):
\begin{equation}
\label{eq:GBM_interactions}
\mathrm{d}h_i = \mu h_i\,\mathrm{d}t + \sigma h_i\,\mathrm{d}W_{i}(t) + k\sum_{i'\in\mathcal{N}(i)}(h_{i'}-h_i)\,\mathrm{d}t,
\end{equation}
where $h_i(t)$ is the building height at site $i$, $t$ is time, $W_i$ is the standard Brownian motion, $k$ is the coupling strength which controls the intensity of local spatial interaction, and $\mathcal{N}(i)$ denotes the neighbours of site $i$. This approach is a spatially embedded and local version of various models previously studied in the literature ~\cite[e.g.,][]{bouchaud2000wealth,bernard2025mean,verbavatz2020growth} and it constitutes a stochastic and parsimonious variant of the reaction-diffusion model~\cite{capel2024angiogenic}.

To evaluate the predictive performance of Eq.~\ref{eq:GBM_interactions}, we performed numerical simulations seeded with the 1993 building‑height spatial distribution (see Methods for details). Across Chinese cities, the simulations reproduce well the observed space and time dynamics (Fig.~\ref{fig:SHE_simulation}) as indicated by the RMSE values. Yet, these results should be interpreted cautiously, because the initial conditions already encode much of the spatial structure, acting as a form of quenched disorder. This is reflected in the small $\sigma$ parameter selected during calibration, which minimises the RMSE between simulations and observations (Supplementary Fig.~\ref{fig:china_all_simu}).

\begin{figure}[h]
    \centering
    \includegraphics[width=1\linewidth]{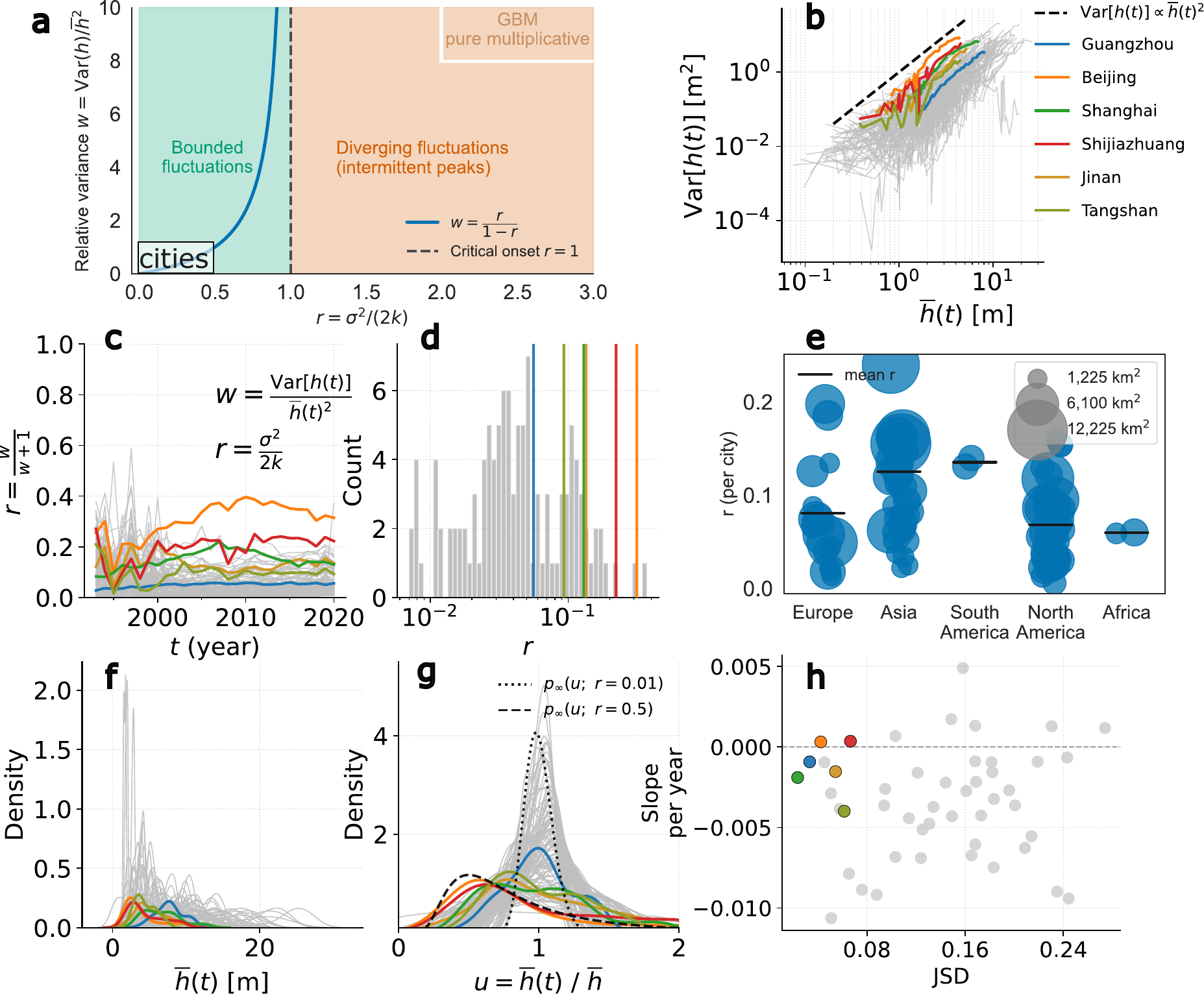}
\caption{\textbf{Growth regimes.} (a) Regime diagram showing the competition between multiplicative noise and mean-field coupling in the stochastic growth model; (b) temporal evolution of variance versus mean; (c) time evolution of the relative fluctuation ratio $r^*(t) = w(t)/(w(t)+1)$; (d) distribution of $r$ from Eq.~\ref{eq:ratio_var_mean} ($r = lim_{t\rightarrow\infty}r^*(t)$ in panel (c)). (e) Empirical value of $r$ computed via Eq.~\ref{eq:ratio_var_mean} for cities worldwide. (f) PDFs of building heights computed from 5km-resolution data for 2020; (g) for the same cities, PDFs of the normalized height $u$, together with the theoretical PDF from Eq.~\ref{eq:stationary_u}; (h) Jensen–Shannon divergence (base 2) between the empirical PDFs of $u$ in the last and previous year along with the annual slope of the JSD relative to the fitted stationary inverse-gamma distribution (negative indicates convergence) for cities larger then 1250$km^2$. Lower-left indicates strongest convergence;  Panels (b–d and f-h) show results for for East Asian cities, cities are represented by a gray line (or a dot in panel h), except for several example cities highlighted in color.} \label{fig:regime_diagram}
\end{figure}

By employing a mean-field approximation (Eq.~\ref{eq:GBM_interactions_MF}; see Methods), we can then characterise distinct growth regimes dependent on the balance between multiplicative noise and local spatial coupling, represented by the dimensionless parameter $ r = \sigma^2/(2k) $ (see Methods). The regime diagram in Fig.~\ref{fig:regime_diagram}a illustrates how existing urban morphologies emerge from the dynamic interplay between noise-driven stochastic fluctuations and local interaction-driven damping. Consistent with this picture, the intra-urban variance of building height scales with the city-average height (Fig.~\ref{fig:regime_diagram}b; Eq.~\ref{eq:ratio_var_mean}). In addition the relative fluctuation $w(t)=\text{Var}[h(t)]/\overline{h}^2(t)$) provides a practical estimator of $r$ in the long-time limit for each city (see Fig.~\ref{fig:regime_diagram}c–e and Methods).

Under the same mean-field approximation (Eq.~\ref{eq:GBM_interactions_MF}), the stationary probability density $p_{\infty}(u)$, of the normalized height $u = h/\overline{h}$ can be derived (see Eq.~\ref{eq:stationary_u} in the Methods section). This stationary law depends only on $r$ (Eq.~\ref{eq:r_parameter}), implying that, in this limit, all city-specific details such as shape or initial conditions become irrelevant. Fig~\ref{fig:regime_diagram}fg shows the empirical distributions of building height for Chinese cities and their normalized counterparts; for reference, curves of $p_{\infty}(u)$ are plotted for two illustrative $r$ values. To assess convergence toward $p_{\infty}(u)$, the annual trend (slope) of the Jensen–Shannon divergence (JSD; \cite{lin2002divergence}) between the empirical PDF of $u$ and $p_{\infty}(u)$ (with $r$ inferred as above) is estimated. Figure~\ref{fig:regime_diagram}h juxtaposes the year 2020 JSD with this annual slope: cities in the lower-left quadrant (small 2020 JSD and negative slope) provide the strongest evidence of convergence toward the theoretical stationary distribution. Notably, large cities such as Guangzhou, Beijing, and Shijiazhuang are already close to stationarity and show little additional convergence over 1993–2020, indicating proximity to the stationary regime from the outset.

\begin{figure}[h]
    \centering
    \includegraphics[width=1\linewidth]{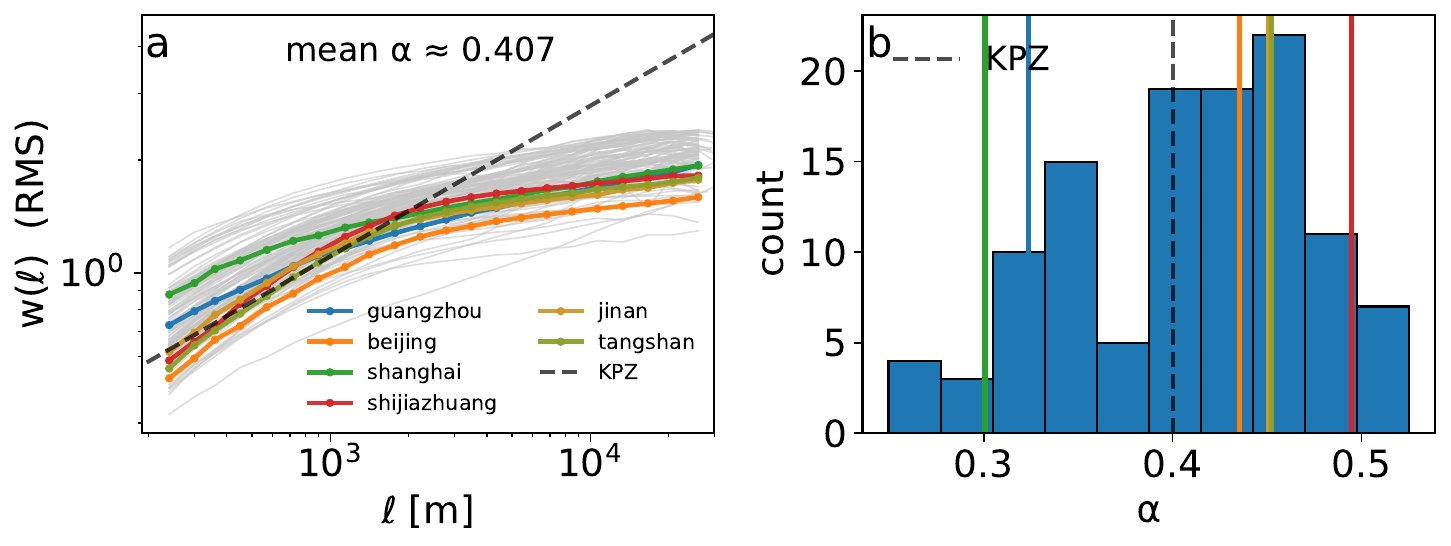}

\caption{\textbf{Multi–scale roughness and KPZ-type scaling across the study cities.} (a) Root-mean-square roughness $w(\ell)=\sqrt{\langle(\phi-\langle\phi\rangle_{\ell})^{2}\rangle_{\ell}}$ versus box size $\ell$ on log--log axes. Thin gray curves show all cities; colored lines highlight examples cities. The roughness exponent $\alpha$ is estimated from the pre-plateau regime. A dashed black reference line (labelled KPZ) indicates $w(\ell)\propto \ell^{0.4}$. \textbf{(b)} Distribution of $\alpha$ across cities; vertical markers denote the highlighted cases. Roughness exponent computed from the data 30-m resoluation data from \cite{chen2025characterizing} for year 2020.}

\label{fig:roughness_multiplot}
\end{figure}

Finally, we bridge our framework for urban growth modelling with the well-established literature on the physics of growing surfaces. Applying the Hopf–Cole transformation to the continuum limit of our model yields the KPZ equation (see Eq.~\ref{eq:KPZ} and Methods), which is a standard model for non-equilibrium interface growth and kinetic roughening \cite{kardar1986dynamic}. We empirically explore this theoretical link using high-resolution urban surface height data from Chinese cities in 2020 \cite{chen2025characterizing}, under the assumption of saturation conditions, consistent with the approach in \cite{najem2020kinetic}. As illustrated in Fig.~\ref{fig:roughness_multiplot}, we computed the width $w(\ell)$ of the Hopf-Cole transformed surface $\phi = \log(h)$ for each cities for different box size $\ell$. We found an empirical roughness scaling $w(\ell)\propto \ell^\alpha$ distributed around $\alpha \approx 0.4$. This result is consistent, for most of the studied cities, with $(2+1)$-dimensional KPZ predictions \cite{pagnani2015numerical}, although temporal constraints in data availability prevent explicit determination of KPZ dynamic exponents.

\section*{Discussion}

Our analysis shows that the vertical growth and internal structure of cities emerge from a delicate balance between multiplicative noise, which amplifies local fluctuations (a process called fluctuation-amplification by Bettencourt \cite{bettencourt2020urban}), and spatial interactions, which diffuse those fluctuations across neighbouring sites (Fig.~\ref{fig:regime_diagram}). This mechanism offers an alternative to the migration terms commonly used in population-exchange models of interacting cities to prevent pure localization \citep{verbavatz2020growth,reia2022spatial,bernard2025mean}, achieving stabilization through fluctuation--diffusion in the physical fabric rather than through exogenous flows.

Once local interactions are coarse-grained to the national scale — treating each city as a single point represented by its time-averaged mean building height — their stabilizing effects largely cancel out and the resulting aggregate trajectory of average building heights is well captured by a pure geometric Brownian motion (Eq.~\ref{eq:GBM}). Strikingly, the same GBM description applies to city GDP, and the mean building height $\overline{h}(t)$ remains tightly correlated with economic output (Fig.~\ref{fig:GBM_country_level}). 

\begin{figure}[h]
    \centering
    \includegraphics[width=1\linewidth]{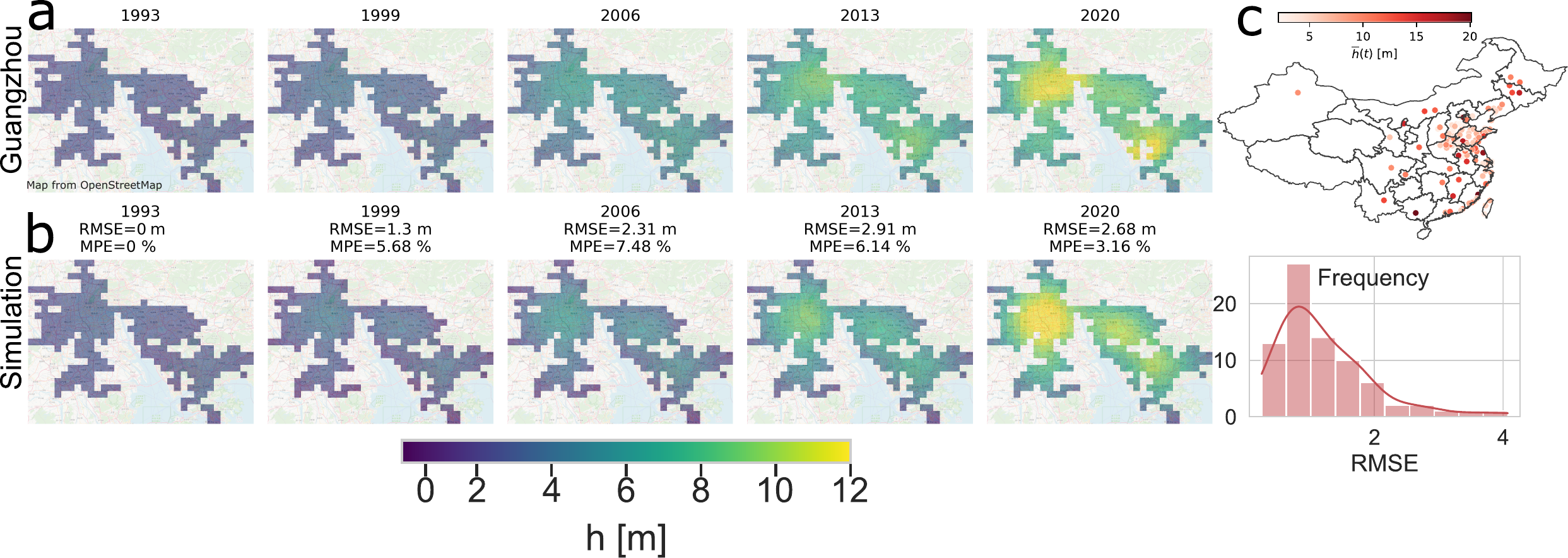}
\caption{\textbf{GBM with local interactions simulation} Results of Eq.~\ref{eq:GBM_interactions}, initialized in 1993 with parameters $(\mu,\sigma,k)$ optimized to minimize the overall RMSE. Guangzhou building height data (a) compared with simulation results (b) at different time steps. (c) Distribution of per-city RMSE values across all Chinese and neighboring cities. The map shows the city locations and their average heights from the data. Panels (a,b) Background maps from OpenStreetMap \cite{OpenStreetMap}.}

    \label{fig:SHE_simulation}
\end{figure}

At finer resolution, our spatially‑extended model (Eq.~\ref{eq:GBM_interactions}) reproduces the observed variance scaling ($\mathrm{Var}[h(t)]\propto \overline{h}^2(t)$; Fig.~\ref{fig:map}b) and morphology of individual Chinese cities (Fig.~\ref{fig:SHE_simulation}), confirming that local interaction terms are essential for describing realistic intra‑urban structures. This multi‑scale construction shows that our framework remains coherent and predictive from neighbourhoods to entire countries (see also SI Figure\ref{fig_si:china_simu_high_res} for a large scale simulation visualization) . This empirical and analytical results provide an explicit example of ``self-similar growth" mechanisms that apply across scale (as introduced by \citet{bettencourt2020urban}).

A practical outcome of these results is the empirical validation of building height as a proxy for the spatial distribution of economic activity. The observed tight $\overline{h}$--GDP coupling (Fig.~\ref{fig:map}g) provides evidence for the stochastic dynamics of economic growth in time-space as proposed by \cite[e.g.][]{gozzi2022stochastic}. It also suggests that remote‑sensing height products can be deployed to monitor sub‑national economic dynamics in data‑scarce regions. Moreover, the dimensionless control parameters $r=\sigma^{2}/(2k)$ and $\gamma=\mu - \sigma^2/2$ (see Methods, Eq.~\ref{eq:r_parameter} and Eq.~\ref{eq:gamma}) provide parsimonious indicators of where a city falls on the spectrum from noise‑dominated intermittent growth to diffusion‑dominated near‑deterministic expansion.

Our model, in the mean-field approximation, admits an analytical expression for the stationary distribution of normalized height (Eq.~\ref{eq:stationary_u}; empirical validation shows in Fig.~\ref{fig:regime_diagram}). This stationary law depends only on $r$ and any dependence on initial conditions vanishes asymptotically. Consequently, all sufficiently mature cities (i.e., those close to stationarity) that share the same $r$ exhibit the same universal stationary distribution. This result provide an example of how universal behaviours emerge from the space–time dynamics of cities, providing a theoretical basis for recent work on urban structure and metabolism \cite[][e.g.]{hendrick2025stochastic,strano2017scaling,la2024urban}, which finds that intra-urban properties collapse onto universal distributions once appropriately rescaled.

In the continuum limit, the developed model (Eq.~\ref{eq:GBM_interactions}) is related to the KPZ equation (Eq.~\ref{eq:KPZ}). This connection motivates measuring the empirical surface roughness $w(\ell)$ and its scaling with the observation window $\ell$ on a 30\,m–resolution dataset \citep{chen2025characterizing} - note that we excluded the 1993--2020 series \citep{frolking2024global} for this analysis because its resolution is too coarse. We have found a power‑law relationship with roughness exponents distributed around the KPZ expected value. These results align with previous works \cite{najem2020kinetic}) showing that cities throughout the Netherlands falls either in the KPZ or the Edwards-Wilkinson universality classes. Although the current four‑epoch 30\,m–resolution dataset \cite{chen2025characterizing} is too sparse to extract the full dynamic Family--Vicsek scaling \cite{vicsek1984dynamic}, the agreement in the static exponent suggests that urban skylines may share the universal roughening statistics of non‑equilibrium growth processes. In the future, annual high‑resolution height maps would enable a definitive test of this conjecture.

Overall, this work anchors the observed dynamic roughening of cities to a general theoretical framework. However, our findings must be interpreted in light of several limitations.  First, our model is most accurate for fast-growing expanding cities that have not yet exhausted their vertical capacity.  Mature metropolises, such as those in the United States, show clear deviations in building-height variance behaviour from geometric Brownian motion (GBM), with an early onset of height dispersion followed by a plateau (SI Fig.\ref{fig:usa_gbm_city}). This behaviour is more naturally captured by the stochastic Fisher-Kolmogorov equation (see Methods; Eq.~\ref{eq:sFK}), which generalizes the heat equation used here with a carrying capacity $K$ and produces a fluctuating saturated state. Second, our empirical analysis is confined largely to East-Asian cities; longer time series (covering period of significant growth) for European and U.S. cities, and the inclusion of additional emerging regions, would be required to confirm the cross‑regional validity of our approach. Third, it would be interesting to also take into account population density; explicitly modelling coupled population--building height dynamics could reveal how horizontal and vertical growth interact and disentangle key differences in residential/workplace occupancy \cite{batista2020uncovering}.

Looking forward, it will be fruitful to generalize the proposed mathematical framework to heterogeneous diffusion constants that reflect transport infrastructure, to incorporate non‑Gaussian construction shocks, and to test whether the building height field indeed satisfies full KPZ universality once extended space-time data become available. Such developments would deepen our understanding of how economic forces, planning regulations, and stochasticity ultimately shape the three‑dimensional structure of modern cities.



\section*{Methods}\label{sec:methods}
\subsection*{Theoretical framework}
We model the evolution of building height $h(t)$ at each spatial location as a stochastic multiplicative process. Initially, we consider heights described by geometric Brownian motion \cite{ross2014variations}:
\begin{equation}
\label{eq:GBM}    
\mathrm{d}h = \mu h\,\mathrm{d}t + \sigma h\,\mathrm{d}W(t),
\end{equation}
where $\mu$ represents deterministic drift linked directly to local GDP growth (Fig.\ref{fig:map}g), $\sigma$ characterizes multiplicative noise amplitude, and $W(t)$ denotes standard Brownian motion. The PDF of $h(t)$ with initial condition $h(0)=h_0$ is the log‐normal distribution
\begin{equation}
\label{eq:GBM_pdf}
  p(h(t))
  = \frac{1}{h(t)\,\sigma\sqrt{2\pi t}}
    \exp\!\left[
      -\frac{\big(\ln(h(t)/h_0) - (\mu-\tfrac{1}{2}\sigma^2)t\big)^2}{2\sigma^2 t}
    \right].
\end{equation} 
For convenience, we define the effective drift
\begin{equation}
\label{eq:gamma}
\gamma \;\equiv\; \mu - \sigma^{2},
\end{equation}
when $\gamma>0$ the drift of $\ln h(t)$ is strictly positive (see e.g. \cite{bettencourt2020urban}), ensuring that every sample path are growing in the long run. This model provides an analytically tractable representation of proportional stochastic growth and is widely used across different domains (from finance \citep{farida2018stock} to hydrology \citep{lefebvre2002geometric} and city science \citep{bettencourt2020urban}).

To introduce realistic spatial structures, we extend the GBM model onto a two-dimensional lattice, incorporating spatial interactions via diffusive coupling between neighbouring sites. The height $ h_i(t) $ at lattice site $ i $ evolves according to:
\begin{equation}
\label{eq:GBM_interactions_meth}
\mathrm{d}h_i = \mu h_i\,\mathrm{d}t + \sigma h_i\,\mathrm{d}W_{i}(t) + k\sum_{i'\in\mathcal{N}(i)}(h_{i'}-h_i)\,\mathrm{d}t,
\end{equation}
where coupling strength $k$ controls the intensity of local spatial interaction, and $\mathcal{N}(i)$ denotes the neighbours of site $i$.

To gain analytical insight, we adopt a mean-field approximation, replacing explicit neighbour interactions with coupling to the mean height $\overline{h}(t)=\mathbb{E}[h(t)]$. Thus, the mean-field dynamics reduce to \cite{bouchaud2000wealth}:
\begin{equation}
\label{eq:GBM_interactions_MF}
\mathrm{d}h = (\mu - k)h\,\mathrm{d}t + k\overline{h}(t)\,\mathrm{d}t + \sigma h\,\mathrm{d}W(t).
\end{equation} Therefore, the mean is (see SI for details):
\begin{equation}
\label{eq:GBM_interactions_mean}
\overline{h}(t)=\overline{h}(0)e^{\mu t}
;\;
\end{equation}
where $\overline{h}(t)$ also corresponds to the mean behaviour of uncoupled GBM. It follows that the variance is (for $k\neq 0$):
\begin{equation}
\label{eq:var_interactions_mean}
\mathrm{Var}[h(t)] = \frac{r}{1 - r}\overline{h}(t)^2 \left( 1 - e^{-2k(1-r)t} \right),
\end{equation}
with $r$ an adimensional
parameter: 
\begin{equation}
\label{eq:r_parameter}
    r=\sigma^2/(2k),
\end{equation}
governing the system behaviour. When $k=0$ we recover the variance of the GBM process:
\begin{equation}
    \mathrm{Var}[h(t)]=\overline{h}(t)^2\bigl(e^{\sigma^2 t}-1\bigr).
    \label{eq:GBM_variance}
\end{equation}

Considering $k>0$, we can identify the following regimes:

\begin{itemize}
\item $r<1$: Coupling dominates noise.  Variance grows proportionally with $\overline{h}^2$, leading to a finite asymptotic relative variance:
\begin{equation}
\label{eq:ratio_var_mean}
w(t):=\frac{\mathrm{Var}[h(t)]
}{\overline{h}^2}\xrightarrow[t\to\infty]{}\frac{r}{1-r}.
\end{equation}
This regime characterizes well the studied cities, see Figs. \ref{fig:map}b and~\ref{fig:regime_diagram}b.
\item $r>1$: Noise dominates coupling (GBM like behaviour). The variance diverges faster than $\overline{h}^2$, causing relative variance to become unbounded, signaling instability dominated by noise fluctuations.
\item $r=1$: Critical case, variance grows as $t\overline{h}(t)^2$.
\end{itemize}
These different regimes are summarized in the diagram reported in Fig.~\ref{fig:regime_diagram}a (finite time and finite size generic simulation illustrating these regimes are provided in the SI Fig.~\ref{fig:regime_illu}). 

The exact Fokker--Planck equation associated with Eq.~\ref{eq:GBM_interactions_MF} is not analytically solvable (this is even more true for the non-approximated model in  Eq.~\ref{eq:GBM_interactions}). However, the normalized variable $u:=h/\overline{h}$ admits a stationary PDF $p_{\infty}$ in the mean-field approximation that is (See Supplementary Information and \cite{bouchaud2000wealth}):
\begin{equation}\label{eq:stationary_u}
    p_\infty(u)
= \frac{(1/r)^{\,1+1/r}}{\Gamma\!\left(1+\tfrac{1}{r}\right)}\,
  u^{-\left(2+\tfrac{1}{r}\right)}\,
  \exp\!\left(-\frac{1}{r\,u}\right).
\end{equation}
Note that the tail of the distribution can be approximated as a Pareto power:
\begin{equation}
P_{\infty}(u)\approx u^{-2 + 1/r}.
\end{equation}
We can express the PDF of $h$ as:
\begin{equation}
p(h(t))=\frac{1}{\overline{h}(t)}\,p\!\left(\frac{h(t)}{\overline{h}(t)}\right).
\end{equation}
If the normalized process \(u(t)\) has reached stationarity, the family \(\{p(h(t))\}_t\) becomes self-similar:
its shape is fixed (given by $p_\infty$) while its scale evolves as \(\overline{h}(t)\).
Before stationarity, however, the shape remains time-dependent,
and self-similarity holds only asymptotically:
\begin{equation}
p(h(t))\xrightarrow[t\to\infty]{}\frac{1}{\overline{h}(t)}\,p_\ast\!\left(\frac{h(t)}{\overline{h}(t)}\right).
\end{equation}
This stationary distribution of the normalized field \(u(t)\) provides
a simple mechanism for the emergence of universal (time-invariant) probability
density functions from a dynamical model, a behaviour reported and observed in various studies
\cite[e.g.][]{verbavatz2020growth,hendrick2025stochastic,la2024urban,strano2017scaling}.

To understand urban morphology at larger scales, we study the continuum limit of the spatially coupled GBM. In this limit, the model follow the stochastic heat equation with multiplicative noise (SHE) \cite{bertini1998two,hu2019some}:
\begin{equation}
\label{eq:SHE}
\partial_t h = D\nabla^2 h + \mu h + \sigma h\eta(\mathbf{x},t),
\end{equation}
where $D=ka^2$ (for small lattice spacing $a\rightarrow 0$) denotes an effective diffusion coefficient and $\eta(\mathbf{x},t)$ is Gaussian white noise. Applying the Hopf–Cole transformation $h(\mathbf{x},t)=\exp[\phi(\mathbf{x},t)]$ \cite{hairer2013solving}, SHE converts to the Kardar–Parisi–Zhang equation \cite{kardar1986dynamic}:
\begin{equation}
\label{eq:KPZ}
\partial_t\phi = D\nabla^2\phi + \frac{\lambda}{2}(\nabla\phi)^2 + \mu + \sigma\eta(\mathbf{x},t),
\end{equation}
with canonical nonlinearity $\lambda=2D$ (the drift term $\mu$ can be absorbed by redefining the field $\phi$ through a Galilean transformation). This explicitly connects our urban growth model to the KPZ universality class, grounding urban spatiotemporal dynamics within a rigorous statistical-physics framework. The KPZ equation was initially developed to describe growth through random deposition and diffusion, it characterizes a universality class thought to include a wide range of models and physical systems \cite{barabasi1995fractal}.

Beyond the stochastic heat equation, the stochastic Fisher-Kolmogorov-Petrovsky-Piscunov (sFKPP; \cite{doering2003interacting}) equation offers a natural generalization for systems that approach a finite carrying capacity $K$.  In two spatial dimensions it reads
\begin{equation}
\label{eq:sFK}
\partial_t h = D \nabla^{2} h + \mu\,h\!\left(1-\frac{h}{K}\right) + \sigma\,h\,\eta(\mathbf{x},t),
\end{equation}
where the logistic term $\mu h(1-h/K)$ introduces a nonlinear saturation that drives the system toward a fluctuating plateau at $h\approx K$.  This regime is conceptually well suited to model mature, land‑limited cities that are close to their maximum vertical capacity, in contrast to rapidly expanding centres described by unbounded stochastic growth (see main text and SI).

\subsection*{Dataset description}

\paragraph{Urban‐height time series (1993–2020).}
We used the global, 1,567 cities dataset of \citet{frolking2024global}, which merges observations from three C‑band Ku‑band radar scatterometers (ERS‑1/2, QuikSCAT, ASCAT) with Landsat‐derived settlement fractions to approximate fractional building cover and volumetric backscatter. Annual composites are provided on a $0.05^{\circ}\!\times\!0.05^{\circ}$ grid ($\approx$ 5km resolution) for 1993–2020, that gives two complementary proxies of vertical development: (i) microwave backscatter, correlated with built volume, and (ii) settlement built‑up fraction, correlated with planimetric expansion. City boundaries follow the Morphological Urban Area (MUA) polygons from \citet{taubenbock2019new}, as adopted to define 1,567 urban areas across 151 countries. Each MUA polygon was overlaid with a 0.05° lat/lon grid and all grid cells contained within or intersecting the polygon were assigned to the city. Grid cells with $\geq$50\% open water were masked using the Global 1-km Consensus Land Cover dataset, and MUAs lacking required inputs were excluded.

\paragraph{High‑resolution anchor heights for China.}
To calibrate and homogenize the radar–optical series over Chinese cities, we employed the 30 m China Multi‑Temporal Built‑up Height product (CMTBH‑30; \cite{chen2025characterizing}). CMTBH‑30 provides per‑pixel mean building height for the benchmark years 2005, 2010, 2015, and 2020, derived with the MTBH‑Net algorithm that uses GEDI lidar returns, Landsat spectral indices, ALOS/PALSAR backscatter, and change metrics (overall $\mathrm{RMSE}<6.5$ m).  

\paragraph{Economic and demographic data.}
Gross Domestic Product (GDP) and population for U.S. metropolitan statistical areas (MSAs) are taken from the Federal Reserve Bank of St.Louis economic database \cite{Fred2025}. Corresponding prefecture‑level GDP and population series for China are obtained from the National Bureau of Statistics of China \cite{NBS2025}.

\subsection*{Data processing}

\paragraph{Inter-satellite bias correction}

The C-/Ku-band backscatter are converted in decibels to unitless power-return ratio \cite{frolking2024global} on the native \(0.05^\circ\) grid using May composites only, remove non-positive values and incomplete metadata, and re-index time so that \(t=0\) is the first observation for each city. Satellites cover partially overlapping periods (ERS 1993–2000, QuikSCAT 1999–2009, ASCAT 2007–2021).

To place all satellites on a common scale we perform a per-pixel intercalibration with QuikSCAT as the reference. For pixels observed by ASCAT and QuikSCAT in 2007–2009, we estimate a pixel-specific affine mapping on the log scale using only the overlap years and apply it to all ASCAT years. Because the ERS–QuikSCAT overlap (1999–2000) is very short, we use a per-pixel intercept-only alignment for ERS, which preserves ERS pixel trends. We refer the reader to the Supplementary Information for further details.


\paragraph{Height conversion in meters (Chinese cities only)}

For China, we convert the power-return ratio to building height using the CMTBH-30 product (30 m; \cite{chen2025characterizing}). We first aggregate CMTBH-30 to the \(0.05^\circ\) analysis grid by averaging all 30 m pixels within each cell, producing a gridded benchmark of building height for 2020 (see also SI Figure \ref{fig_si:30m_vs_frokling}). For each city, we relate the 2020 May power-return ratio to the aggregated 2020 building height across all grid cells with both datasets available, using a simple city-specific linear calibration. The fitted city-level coefficients are then applied to the full time series of May power-return ratios to obtain annual building heights in meters for every year in the study period.

For cities outside China, we retain the reconstructed May power-return series $h^{\text{May}}_{c,i,t}$ as the vertical-development proxy without converting to meters. This is appropriate because our GBM-based inference targets relative temporal change (log-drift $\gamma$) and volatility $\sigma$, both of which are invariant to multiplicative rescaling of $h$.

\subsection*{Numerical simulations and calibration}

We simulate Eq.~\eqref{eq:GBM_interactions} on a discrete $2$D grid of cells whose values represent building heights. We used the Euler--Maruyama scheme \cite{kloeden2012numerical}:
\begin{equation}
h_i^{t+\Delta t}
= h_i^{t}
+ \mu\, h_i^{t}\,\Delta t
+ \sigma\, h_i^{t}\,\sqrt{\Delta t}\,\xi_i^{t}
+ k\,\Delta t \sum_{j\in\mathcal{N}(i)} \big(h_j^{t}-h_i^{t}\big).
\label{eq:EM_update}
\end{equation}
with i.i.d.\ standard normal shocks $\xi_i^t\sim\mathcal{N}(0,1)$ and
$\mathcal{N}(i)$ is the 8-nearest neighbourhood. Heights are initialized from the first observation year (1993), $h_i^{t_0}=H_i^{\mathrm{obs}}(1993)$.

Cities are simulated on fixed spatial domains and cells outside the mapped urban extent are treated as inactive via a boolean mask. These cells are thus excluded from neighborhood sums in Eq.~\eqref{eq:EM_update} and all loss calculations; this is equivalent to imposing no-flux boundary conditions along the domain perimeter.

For each city, parameters $(\mu,\sigma,k)$ are estimated by minimizing the mean RMSE between simulations and observations across all available years. For any candidate $(\mu,\sigma,k)$ we run $R$ stochastic replicas and compare the ensemble mean $\bar h_i^t=\frac{1}{R}\sum_{r=1}^{R} h_{i}^{t,(r)}$ to the data over the active set $\Omega$:
\begin{equation}
\mathrm{RMSE}_t
= \left(\frac{1}{|\Omega|}\sum_{i\in\Omega}\Big(\bar h_i^t - H_i^{\mathrm{obs}}(t)\Big)^2\right)^{1/2},
\qquad
\mathcal{L}(\mu,\sigma,k)
= \frac{1}{T}\sum_{t=1}^{T} \mathrm{RMSE}_t.
\end{equation}
We minimize $\mathcal{L}$ subject to $\sigma\ge 0$ and $k\ge 0$ using a coarse grid search followed by a local derivative-free optimizer based on a quasi-Newton methods called Limited-memory BFGS \cite{liu1989limited}. Error metrics are computed only over active (non-masked) cells.

\section*{Author Contributions}
G.M. and M.H. conceived the study. M.H. performed the analyses and drafted the manuscript. M.H. and G.M interpreted the results, provided feedback that helped shape the analysis, and contributed to writing the manuscript.

\section*{Competing Interests}
The authors declare no competing interests.

\section*{Data Availability}

All data used in this article are freely accessible and referenced in the text.

\bibliographystyle{unsrtnat} 

\bibliography{ref}

\begin{thebibliography}{56}
\providecommand{\natexlab}[1]{#1}
\providecommand{\url}[1]{\texttt{#1}}
\expandafter\ifx\csname urlstyle\endcsname\relax
  \providecommand{\doi}[1]{doi: #1}\else
  \providecommand{\doi}{doi: \begingroup \urlstyle{rm}\Url}\fi

\bibitem[Liu et~al.(2020)Liu, Huang, Xu, Li, Li, Ciais, Lin, Gong, Ziegler, Chen, et~al.]{liu2020high}
Xiaoping Liu, Yinghuai Huang, Xiaocong Xu, Xuecao Li, Xia Li, Philippe Ciais, Peirong Lin, Kai Gong, Alan~D Ziegler, Anping Chen, et~al.
\newblock High-spatiotemporal-resolution mapping of global urban change from 1985 to 2015.
\newblock \emph{Nature Sustainability}, 3\penalty0 (7):\penalty0 564--570, 2020.

\bibitem[Grimm et~al.(2008)Grimm, Faeth, Golubiewski, Redman, Wu, Bai, and Briggs]{grimm2008global}
Nancy~B Grimm, Stanley~H Faeth, Nancy~E Golubiewski, Charles~L Redman, Jianguo Wu, Xuemei Bai, and John~M Briggs.
\newblock Global change and the ecology of cities.
\newblock \emph{science}, 319\penalty0 (5864):\penalty0 756--760, 2008.

\bibitem[Bettencourt et~al.(2007)Bettencourt, Lobo, Helbing, K{\"u}hnert, and West]{bettencourt2007growth}
Lu{\'\i}s~MA Bettencourt, Jos{\'e} Lobo, Dirk Helbing, Christian K{\"u}hnert, and Geoffrey~B West.
\newblock Growth, innovation, scaling, and the pace of life in cities.
\newblock \emph{Proceedings of the national academy of sciences}, 104\penalty0 (17):\penalty0 7301--7306, 2007.

\bibitem[Verbavatz and Barthelemy(2020)]{verbavatz2020growth}
Vincent Verbavatz and Marc Barthelemy.
\newblock The growth equation of cities.
\newblock \emph{Nature}, 587\penalty0 (7834):\penalty0 397--401, 2020.

\bibitem[Barthelemy(2019)]{barthelemy2019statistical}
Marc Barthelemy.
\newblock The statistical physics of cities.
\newblock \emph{Nature Reviews Physics}, 1\penalty0 (6):\penalty0 406--415, 2019.

\bibitem[Gabaix(1999)]{gabaix1999zipf}
Xavier Gabaix.
\newblock Zipf's law for cities: an explanation.
\newblock \emph{The Quarterly journal of economics}, 114\penalty0 (3):\penalty0 739--767, 1999.

\bibitem[Rozenfeld et~al.(2011)Rozenfeld, Rybski, Gabaix, and Makse]{rozenfeld2011area}
Hern{\'a}n~D Rozenfeld, Diego Rybski, Xavier Gabaix, and Hern{\'a}n~A Makse.
\newblock The area and population of cities: New insights from a different perspective on cities.
\newblock \emph{American Economic Review}, 101\penalty0 (5):\penalty0 2205--2225, 2011.

\bibitem[Alonso(1964)]{alonso1964location}
William Alonso.
\newblock \emph{Location and land use: Toward a general theory of land rent}.
\newblock Harvard university press, 1964.

\bibitem[Fischer(2011)]{fischer2011central}
Kathrin Fischer.
\newblock Central places: The theories of von th{\"u}nen, christaller, and l{\"o}sch.
\newblock In \emph{Foundations of location analysis}, pages 471--505. Springer, 2011.

\bibitem[Krugman(1996)]{krugman1996self}
Paul Krugman.
\newblock \emph{The self organizing economy}.
\newblock John Wiley \& Sons, 1996.

\bibitem[Batty and Longley(1986)]{batty1986fractal}
Michael Batty and Paul~A Longley.
\newblock The fractal simulation of urban structure.
\newblock \emph{Environment and Planning a}, 18\penalty0 (9):\penalty0 1143--1179, 1986.

\bibitem[Frankhauser(1994)]{frankhauser1994fractalite}
Pierre Frankhauser.
\newblock \emph{La fractalit{\'e} des structures urbaines}.
\newblock Anthropos, 1994.

\bibitem[Batty(1991)]{batty1991cities}
Michael Batty.
\newblock Cities as fractals: simulating growth and form.
\newblock In \emph{Fractals and chaos}, pages 43--69. Springer, 1991.

\bibitem[Makse et~al.(1998)Makse, Andrade, Batty, Havlin, Stanley, et~al.]{makse1998modeling}
Hern{\'a}n~A Makse, Jos{\'e}~S Andrade, Michael Batty, Shlomo Havlin, H~Eugene Stanley, et~al.
\newblock Modeling urban growth patterns with correlated percolation.
\newblock \emph{Physical Review E}, 58\penalty0 (6):\penalty0 7054, 1998.

\bibitem[Raimbault(2018)]{raimbault2018calibration}
Juste Raimbault.
\newblock Calibration of a density-based model of urban morphogenesis.
\newblock \emph{PloS one}, 13\penalty0 (9):\penalty0 e0203516, 2018.

\bibitem[Tirico et~al.(2018)Tirico, Balev, Dutot, and Olivier]{tirico2018morphogenesis}
Michele Tirico, Stefan Balev, Antoine Dutot, and Damien Olivier.
\newblock Morphogenesis of complex networks: a reaction diffusion framework for spatial graphs.
\newblock In \emph{International Conference on Complex Networks and their Applications}, pages 769--781. Springer, 2018.

\bibitem[Capel-Timms et~al.(2024)Capel-Timms, Levinson, Lahoorpoor, Bonetti, and Manoli]{capel2024angiogenic}
Isabella Capel-Timms, David Levinson, Bahman Lahoorpoor, Sara Bonetti, and Gabriele Manoli.
\newblock The angiogenic growth of cities.
\newblock \emph{Journal of the Royal Society Interface}, 21\penalty0 (213):\penalty0 20230657, 2024.

\bibitem[Najem et~al.(2020)Najem, Krayem, Ala-Nissila, and Grant]{najem2020kinetic}
Sara Najem, Alaa Krayem, Tapio Ala-Nissila, and Martin Grant.
\newblock Kinetic roughening of the urban skyline.
\newblock \emph{Physical Review E}, 101\penalty0 (5):\penalty0 050301, 2020.

\bibitem[Marquis et~al.(2025)Marquis, Artime, Gallotti, and Barthelemy]{marquis2025universal}
Ulysse Marquis, Oriol Artime, Riccardo Gallotti, and Marc Barthelemy.
\newblock Universal roughness and the dynamics of urban expansion.
\newblock \emph{arXiv preprint arXiv:2506.10656}, 2025.

\bibitem[Reia et~al.(2022)Reia, Rao, Barthelemy, and Ukkusuri]{reia2022spatial}
Sandro~M Reia, P~Suresh~C Rao, Marc Barthelemy, and Satish~V Ukkusuri.
\newblock Spatial structure of city population growth.
\newblock \emph{Nature communications}, 13\penalty0 (1):\penalty0 5931, 2022.

\bibitem[Jin et~al.(2023)Jin, Wang, Ge, and Yan]{jin2023detecting}
Min Jin, Lizhe Wang, Fudong Ge, and Jining Yan.
\newblock Detecting the interaction between urban elements evolution with population dynamics model.
\newblock \emph{Scientific Reports}, 13\penalty0 (1):\penalty0 12367, 2023.

\bibitem[Gibrat(1931)]{gibrat1931inegalites}
R~Gibrat.
\newblock Les in{\'e}galit{\'e}s {\'e}conomiques, librairie du recueil sirey, paris.
\newblock \emph{What makes firms grow in developing countries}, 169, 1931.

\bibitem[Eeckhout(2004)]{eeckhout2004gibrat}
Jan Eeckhout.
\newblock Gibrat's law for (all) cities.
\newblock \emph{American Economic Review}, 94\penalty0 (5):\penalty0 1429--1451, 2004.

\bibitem[Bettencourt(2020)]{bettencourt2020urban}
Luis~MA Bettencourt.
\newblock Urban growth and the emergent statistics of cities.
\newblock \emph{Science advances}, 6\penalty0 (34):\penalty0 eaat8812, 2020.

\bibitem[Bonetti and Porporato(2017)]{bonetti2017dynamic}
Sara Bonetti and A~Porporato.
\newblock On the dynamic smoothing of mountains.
\newblock \emph{Geophysical Research Letters}, 44\penalty0 (11):\penalty0 5531--5539, 2017.

\bibitem[Munoz(2003)]{munoz2003multiplicativenoisenonequilibriumphase}
Miguel~A. Munoz.
\newblock Multiplicative noise in non-equilibrium phase transitions: A tutorial, 2003.
\newblock URL \url{https://arxiv.org/abs/cond-mat/0303650}.

\bibitem[Bouchaud and M{\'e}zard(2000)]{bouchaud2000wealth}
Jean-Philippe Bouchaud and Marc M{\'e}zard.
\newblock Wealth condensation in a simple model of economy.
\newblock \emph{Physica A: Statistical Mechanics and its Applications}, 282\penalty0 (3-4):\penalty0 536--545, 2000.

\bibitem[Bernard et~al.(2025)Bernard, Bouchaud, and Doussal]{bernard2025mean}
Maximilien Bernard, Jean-Philippe Bouchaud, and Pierre~Le Doussal.
\newblock A mean-field theory for heterogeneous random growth with redistribution.
\newblock \emph{arXiv preprint arXiv:2503.23189}, 2025.

\bibitem[Barr and Luo(2021)]{barr2021growing}
Jason Barr and Jingshu Luo.
\newblock Growing skylines: The economic determinants of skyscrapers in china.
\newblock \emph{The journal of real estate finance and economics}, 63:\penalty0 210--248, 2021.

\bibitem[Frolking et~al.(2024)Frolking, Mahtta, Milliman, Esch, and Seto]{frolking2024global}
Steve Frolking, Richa Mahtta, Tom Milliman, Thomas Esch, and Karen~C Seto.
\newblock Global urban structural growth shows a profound shift from spreading out to building up.
\newblock \emph{Nature Cities}, pages 1--12, 2024.

\bibitem[Kardar et~al.(1986)Kardar, Parisi, and Zhang]{kardar1986dynamic}
Mehran Kardar, Giorgio Parisi, and Yi-Cheng Zhang.
\newblock Dynamic scaling of growing interfaces.
\newblock \emph{Physical Review Letters}, 56\penalty0 (9):\penalty0 889, 1986.

\bibitem[Wakita et~al.(1997)Wakita, Itoh, Matsuyama, and Matsushita]{wakita1997self}
Jun-ichi Wakita, Hiroto Itoh, Tohey Matsuyama, and Mitsugu Matsushita.
\newblock Self-affinity for the growing interface of bacterial colonies.
\newblock \emph{Journal of the Physical Society of Japan}, 66\penalty0 (1):\penalty0 67--72, 1997.

\bibitem[Lin(2002)]{lin2002divergence}
Jianhua Lin.
\newblock Divergence measures based on the shannon entropy.
\newblock \emph{IEEE Transactions on Information theory}, 37\penalty0 (1):\penalty0 145--151, 2002.

\bibitem[Chen et~al.(2025)Chen, Huang, Qin, Liu, Wu, Zhao, Liu, Wang, Li, Cheng, et~al.]{chen2025characterizing}
Peimin Chen, Huabing Huang, Peng Qin, Xiangjiang Liu, Zhenbang Wu, Feng Zhao, Chong Liu, Jie Wang, Zhan Li, Xiao Cheng, et~al.
\newblock Characterizing dynamics of built-up height in china from 2005 to 2020 based on gedi, landsat, and palsar data.
\newblock \emph{Remote Sensing of Environment}, 325:\penalty0 114776, 2025.

\bibitem[Pagnani and Parisi(2015)]{pagnani2015numerical}
Andrea Pagnani and Giorgio Parisi.
\newblock Numerical estimate of the kardar-parisi-zhang universality class in (2+ 1) dimensions.
\newblock \emph{Physical Review E}, 92\penalty0 (1):\penalty0 010101, 2015.

\bibitem[Gozzi and Leocata(2022)]{gozzi2022stochastic}
Fausto Gozzi and Marta Leocata.
\newblock A stochastic model of economic growth in time-space.
\newblock \emph{SIAM Journal on Control and Optimization}, 60\penalty0 (2):\penalty0 620--651, 2022.

\bibitem[Hendrick et~al.(2025)Hendrick, Rinaldo, and Manoli]{hendrick2025stochastic}
Martin Hendrick, Andrea Rinaldo, and Gabriele Manoli.
\newblock A stochastic theory of urban metabolism.
\newblock \emph{Proceedings of the National Academy of Sciences}, 122\penalty0 (33):\penalty0 e2501224122, 2025.

\bibitem[Strano et~al.(2017)Strano, Giometto, Shai, Bertuzzo, Mucha, and Rinaldo]{strano2017scaling}
Emanuele Strano, Andrea Giometto, Saray Shai, Enrico Bertuzzo, Peter~J Mucha, and Andrea Rinaldo.
\newblock The scaling structure of the global road network.
\newblock \emph{Royal Society open science}, 4\penalty0 (10):\penalty0 170590, 2017.

\bibitem[La~Porta and Zapperi(2024)]{la2024urban}
Caterina~AM La~Porta and Stefano Zapperi.
\newblock Urban scaling functions: Emission, pollution and health.
\newblock \emph{Journal of Urban Health}, 101\penalty0 (4):\penalty0 752--763, 2024.

\bibitem[Vicsek and Family(1984)]{vicsek1984dynamic}
Tam{\'a}s Vicsek and Fereydoon Family.
\newblock Dynamic scaling for aggregation of clusters.
\newblock \emph{Physical Review Letters}, 52\penalty0 (19):\penalty0 1669, 1984.

\bibitem[Batista~e Silva et~al.(2020)Batista~e Silva, Freire, Schiavina, Rosina, Mar{\'\i}n-Herrera, Ziemba, Craglia, Koomen, and Lavalle]{batista2020uncovering}
Filipe Batista~e Silva, S{\'e}rgio Freire, Marcello Schiavina, Kon{\v{s}}tant{\'\i}n Rosina, Mario~Alberto Mar{\'\i}n-Herrera, Lukasz Ziemba, Massimo Craglia, Eric Koomen, and Carlo Lavalle.
\newblock Uncovering temporal changes in europe’s population density patterns using a data fusion approach.
\newblock \emph{Nature communications}, 11\penalty0 (1):\penalty0 4631, 2020.

\bibitem[Ross(2014)]{ross2014variations}
Sheldon~M Ross.
\newblock Variations on brownian motion.
\newblock \emph{Introduction to Probability Models}, pages 612--614, 2014.

\bibitem[Farida~Agustini et~al.(2018)Farida~Agustini, Affianti, and Putri]{farida2018stock}
W~Farida~Agustini, Ika~Restu Affianti, and Endah~RM Putri.
\newblock Stock price prediction using geometric brownian motion.
\newblock In \emph{Journal of physics: conference series}, volume 974, page 012047. IOP Publishing, 2018.

\bibitem[Lefebvre(2002)]{lefebvre2002geometric}
Mario Lefebvre.
\newblock Geometric brownian motion as a model for river flows.
\newblock \emph{Hydrological processes}, 16\penalty0 (7):\penalty0 1373--1381, 2002.

\bibitem[Bertini and Cancrini(1998)]{bertini1998two}
Lorenzo Bertini and Nicoletta Cancrini.
\newblock The two-dimensional stochastic heat equation: renormalizing a multiplicative noise.
\newblock \emph{Journal of Physics A: Mathematical and General}, 31\penalty0 (2):\penalty0 615, 1998.

\bibitem[Hu(2019)]{hu2019some}
Yaozhong Hu.
\newblock Some recent progress on stochastic heat equations.
\newblock \emph{Acta Mathematica Scientia}, 39\penalty0 (3):\penalty0 874--914, 2019.

\bibitem[Hairer(2013)]{hairer2013solving}
Martin Hairer.
\newblock Solving the kpz equation.
\newblock \emph{Annals of mathematics}, pages 559--664, 2013.

\bibitem[Barab{\'a}si and Stanley(1995)]{barabasi1995fractal}
A-L Barab{\'a}si and Harry~Eugene Stanley.
\newblock \emph{Fractal concepts in surface growth}.
\newblock Cambridge university press, 1995.

\bibitem[Doering et~al.(2003)Doering, Mueller, and Smereka]{doering2003interacting}
Charles~R Doering, Carl Mueller, and Peter Smereka.
\newblock Interacting particles, the stochastic fisher--kolmogorov--petrovsky--piscounov equation, and duality.
\newblock \emph{Physica A: Statistical Mechanics and its Applications}, 325\penalty0 (1-2):\penalty0 243--259, 2003.

\bibitem[Taubenb{\"o}ck et~al.(2019)Taubenb{\"o}ck, Weigand, Esch, Staab, Wurm, Mast, and Dech]{taubenbock2019new}
Hannes Taubenb{\"o}ck, Matthias Weigand, Thomas Esch, Jeroen Staab, Michael Wurm, Johannes Mast, and Stefan Dech.
\newblock A new ranking of the world's largest cities—do administrative units obscure morphological realities?, 2019.

\bibitem[Fre(2025)]{Fred2025}
Federal reserve economic data.
\newblock \url{https://fred.stlouisfed.org/}, 2025.
\newblock Accessed: 2025-07-28.

\bibitem[NBS(2025)]{NBS2025}
National bureau of statistics of china.
\newblock \url{https://www.stats.gov.cn/english/}, 2025.
\newblock Accessed: 2025-07-28.

\bibitem[Kloeden et~al.(2012)Kloeden, Platen, and Schurz]{kloeden2012numerical}
Peter~Eris Kloeden, Eckhard Platen, and Henri Schurz.
\newblock \emph{Numerical solution of SDE through computer experiments}.
\newblock Springer Science \& Business Media, 2012.

\bibitem[Liu and Nocedal(1989)]{liu1989limited}
Dong~C Liu and Jorge Nocedal.
\newblock On the limited memory bfgs method for large scale optimization.
\newblock \emph{Mathematical programming}, 45\penalty0 (1):\penalty0 503--528, 1989.

\bibitem[{OpenStreetMap contributors}(2017)]{OpenStreetMap}
{OpenStreetMap contributors}.
\newblock {Planet dump retrieved from https://planet.osm.org }.
\newblock \url{ https://www.openstreetmap.org }, 2017.

\bibitem[Freedman and Diaconis(1981)]{freedman1981histogram}
David Freedman and Persi Diaconis.
\newblock On the histogram as a density estimator: L 2 theory.
\newblock \emph{Zeitschrift f{\"u}r Wahrscheinlichkeitstheorie und verwandte Gebiete}, 57\penalty0 (4):\penalty0 453--476, 1981.

\end{thebibliography}
\clearpage 
\section*{Figures}

\clearpage 
\appendix
\renewcommand{\thefigure}{S\arabic{figure}}
\renewcommand{\theequation}{S\arabic{equation}}
\setcounter{figure}{0}
\setcounter{equation}{0}
\section*{Supplementary Information}
\subsection*{Supplementary Figures}
\begin{figure}[h!]
    \centering
    \includegraphics[width=1\linewidth]{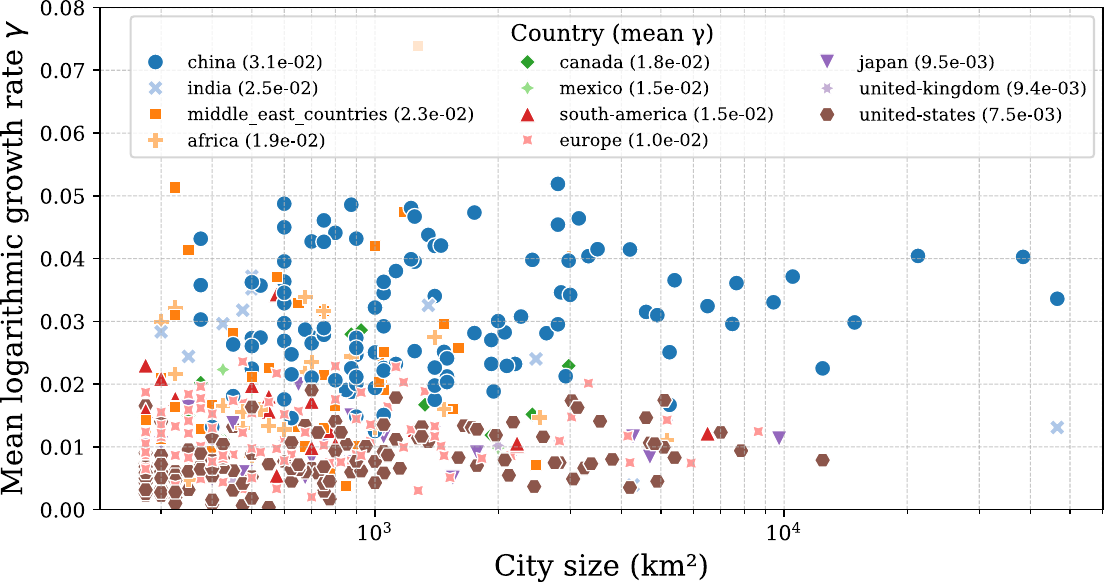}
\caption{Mean logarithmic growth rate $\gamma$ (Eq.~\ref{eq:gamma}) as a function of city size across all cities in the sample; points are colored by country/region.}
\label{fig:gamma_size}
\end{figure}


\begin{figure}[h]
    \centering
    \includegraphics[width=1\linewidth]{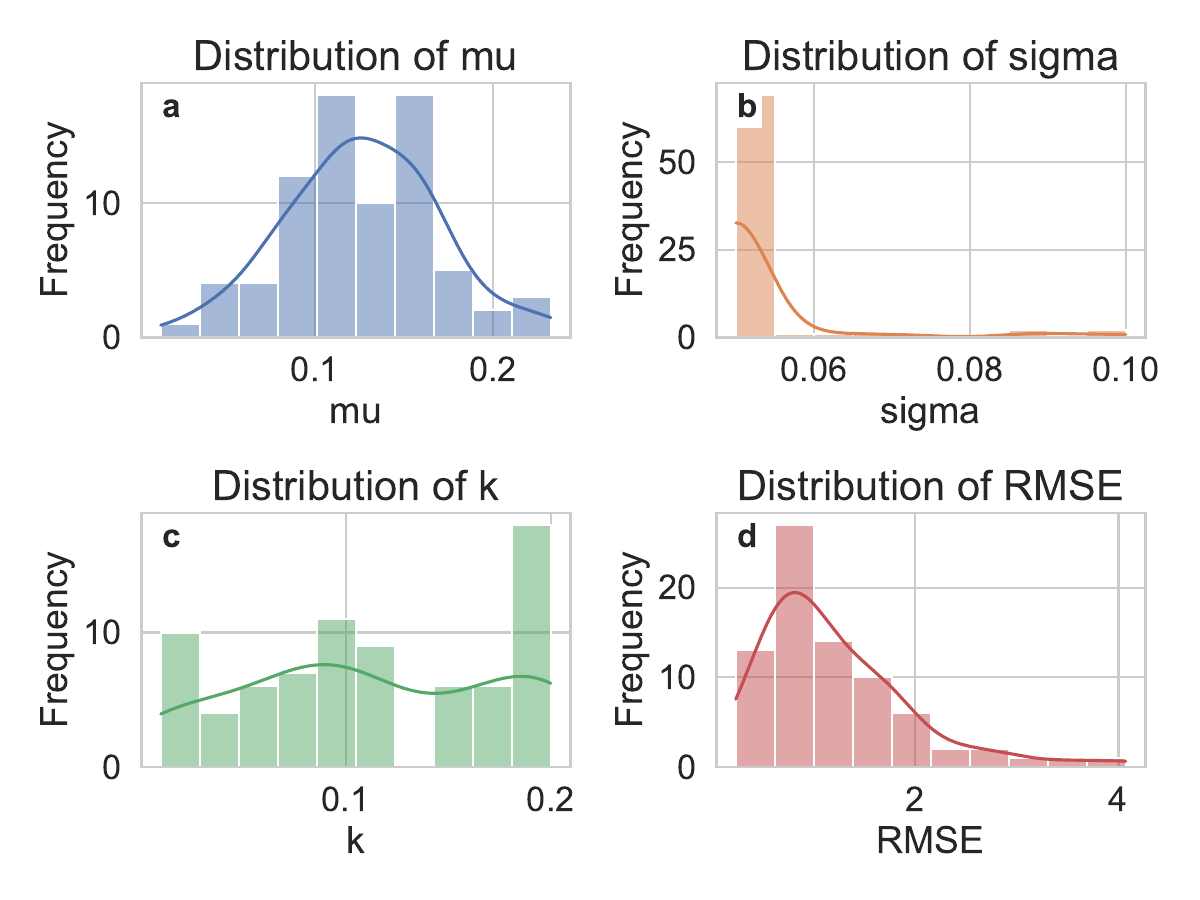}
    \caption{Distributions of the calibrated parameters and simulation error across Chinese cities: (a) growth rate $\mu$ (b) volatility $\sigma$(c) interaction strength $k$, and (d) RMSE between simulated and observed trajectories.}
    \label{fig:china_all_simu}
\end{figure}

\begin{figure}[h]
    \centering
    \includegraphics[width=1\linewidth]{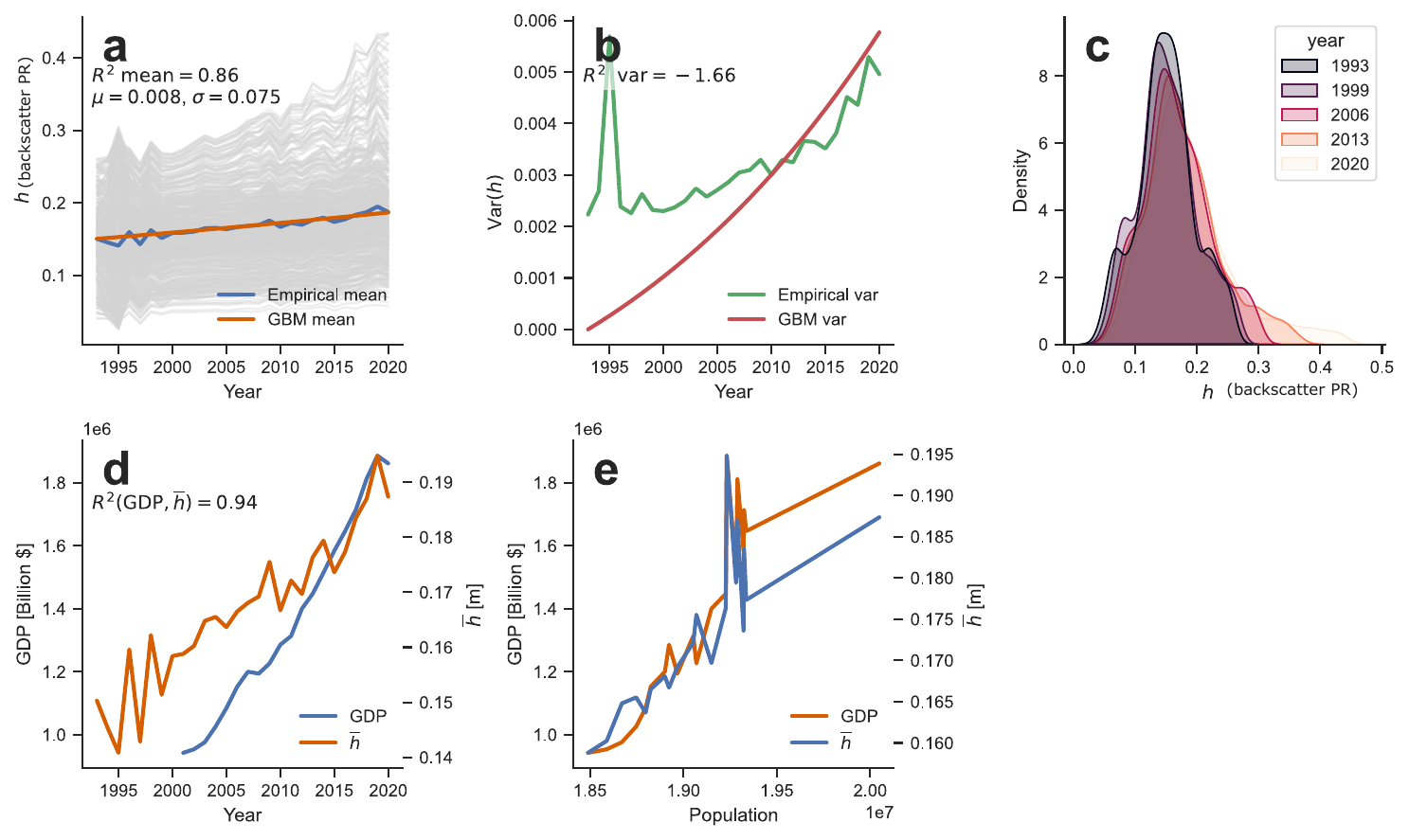}
\caption{\textbf{New-York city:} Comparison of empirical building–height statistics with the geometric Brownian‑motion (GBM) model for each city.  
\textbf{(a)} Mean of pixel height $h(t)$; \textbf{(b)} height variance $\mathrm{Var}(h)$; \textbf{(c)} kernel‑density estimates of height for five benchmark years; \textbf{(d)} co‑evolution of GDP and mean height; \textbf{(e)} GDP versus population with height overlay. Shaded grey lines show pixel buildings height, coloured lines show city means (blue) and GBM predictions (orange/red). In‑panel boxes list fitted GBM parameters and the corresponding coefficients of determination $R^{2}$.}
    \label{fig:NY}
\end{figure}
\begin{figure}[h]
    \centering
    \includegraphics[width=1\linewidth]{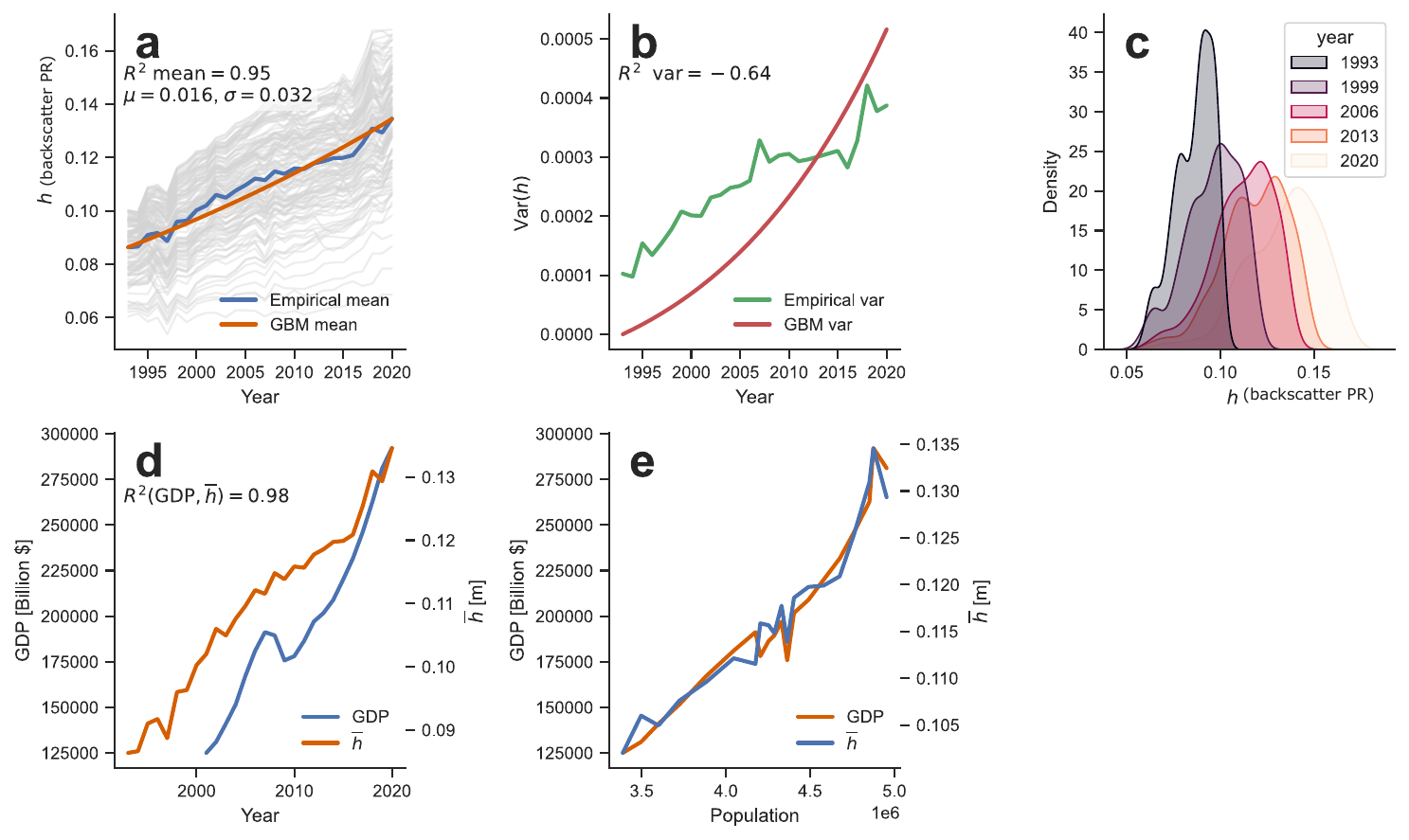}
    \caption{\textbf{Phoenix:} Comparison of empirical building–height statistics with the geometric Brownian‑motion (GBM) model for each city.\textbf{(a)} Mean height $h(t)$; \textbf{(b)} height variance $\mathrm{Var}(h)$; \textbf{(c)} kernel‑density estimates of height for five benchmark years; \textbf{(d)} co‑evolution of GDP and mean height; \textbf{(e)} GDP versus population with height overlay; \textbf{(f)} population‑based view of GDP–height coupling.  Shaded grey lines show pixel buildings height, coloured lines show city means (blue) and GBM predictions (orange/red).  In‑panel boxes list fitted GBM parameters and the corresponding coefficients of determination $R^{2}$.}
    \label{fig:phoenix}
\end{figure}
\begin{figure}[h]
    \centering
    \includegraphics[width=1\linewidth]{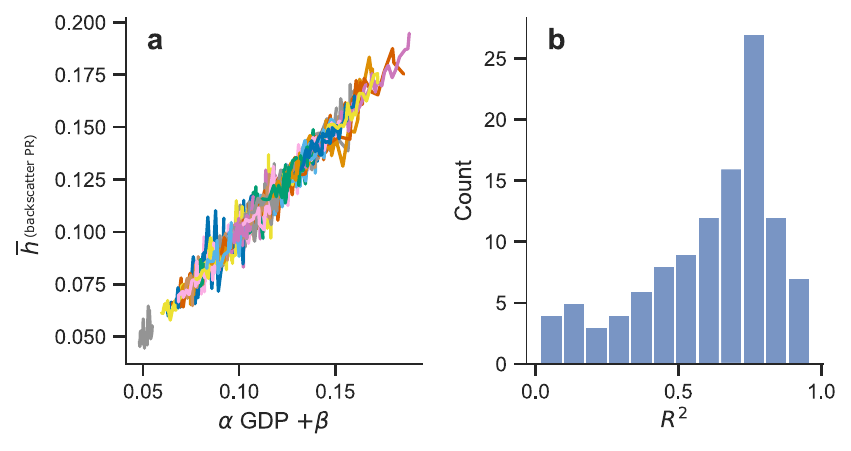}
    \caption{\textbf{(a)} Mean USA city height express as linear linear combination of city GDP.\textbf{(a)} Coefficient of determination per city.}
    \label{fig:gdp_vs_hieght_usa}
\end{figure}

\begin{figure}[h]
    \centering
    \includegraphics[width=1\linewidth]{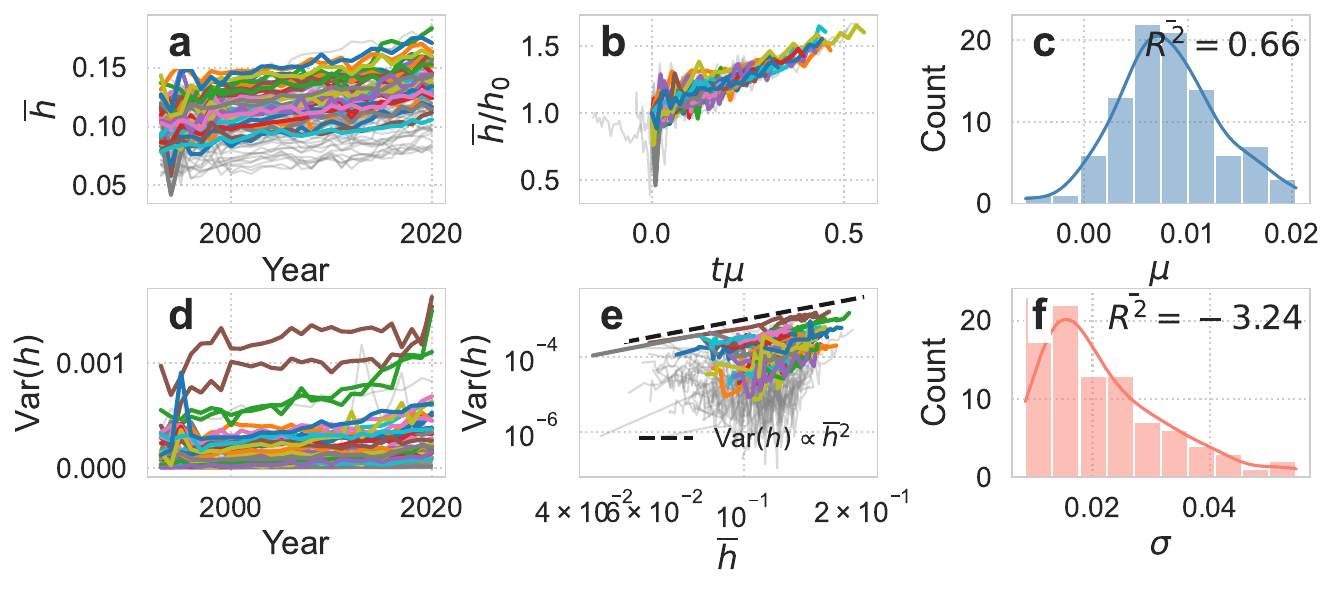}
    \caption{\textbf{Largest USA Cities:}
(a) Empirical mean building height $\overline{h}(t)$ for each city. 
(b) Same trajectories rescaled by their drift term ($t\mu$). 
(c) Distribution of estimated drift coefficients. 
(d) Empirical variance of building heights over time. 
(e) Relationship between variance and mean height. 
(f) Distribution of estimated volatility values.}
    \label{fig:usa_gbm_city}
\end{figure}




\begin{figure}[h]
    \centering
    \includegraphics[width=1\linewidth]{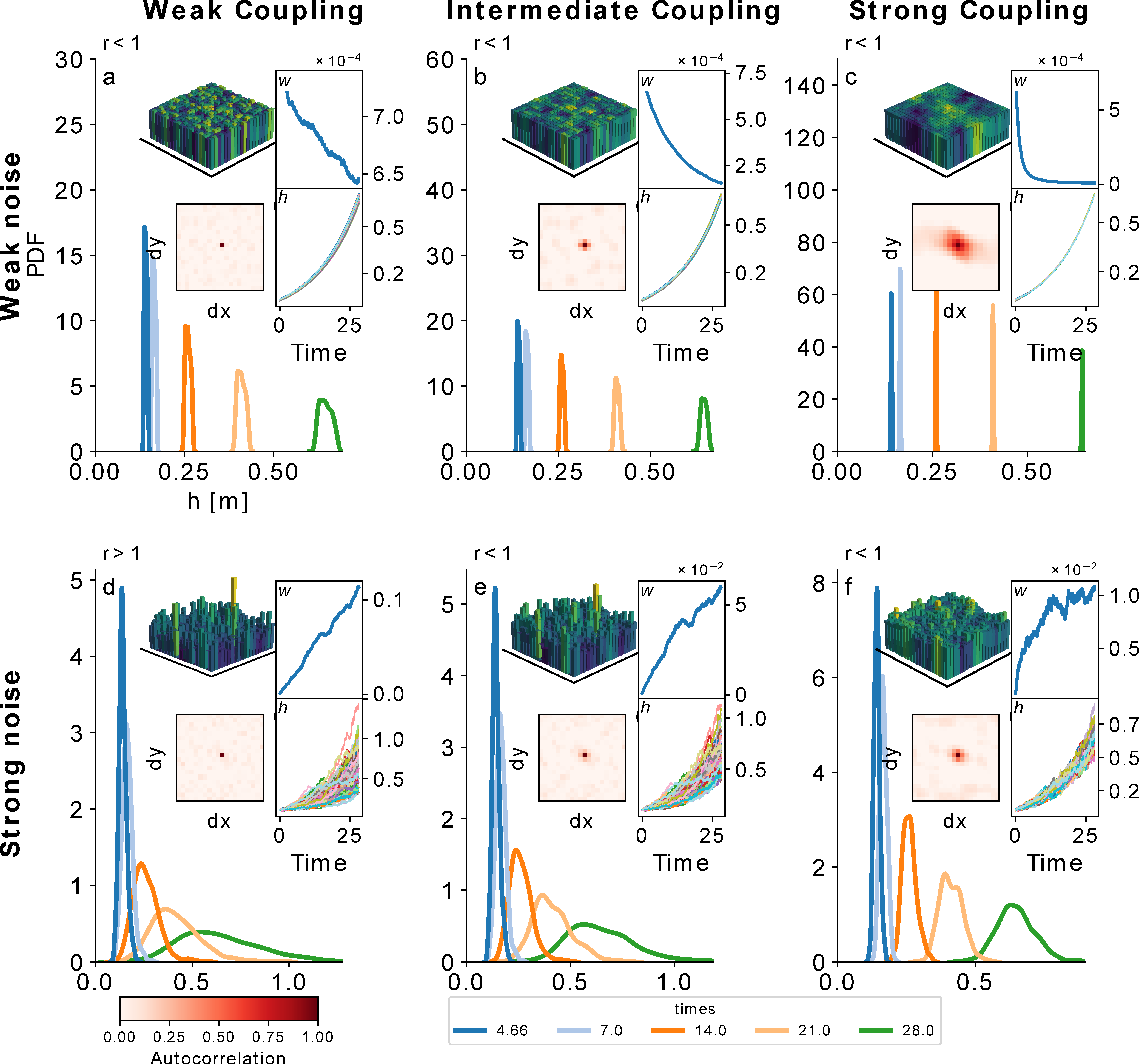}
\caption{%
Synthetic height fields generated by the stochastic growth model (Eq.\ref{eq:GBM_interactions})
for six parameter regimes (weak noise (a-c),
strong noise (d-f); Main panels: kernel–density estimates of the height probability density
at four times. Insets (per panel, top$\rightarrow$bottom/left$\rightarrow$right) show: final 3‑D height surface,
$20\!\times\!20$‑pixel autocorrelation of the same surface,
time series of the relative variance $w(t)=\mathrm{Var}(h)/\overline{h}^{2}$, and trajectories of 100 randomly sampled pixels $h(t)$.}
\label{fig:regime_illu}
\end{figure}

\begin{figure}
    \centering
    \includegraphics[width=1\linewidth]{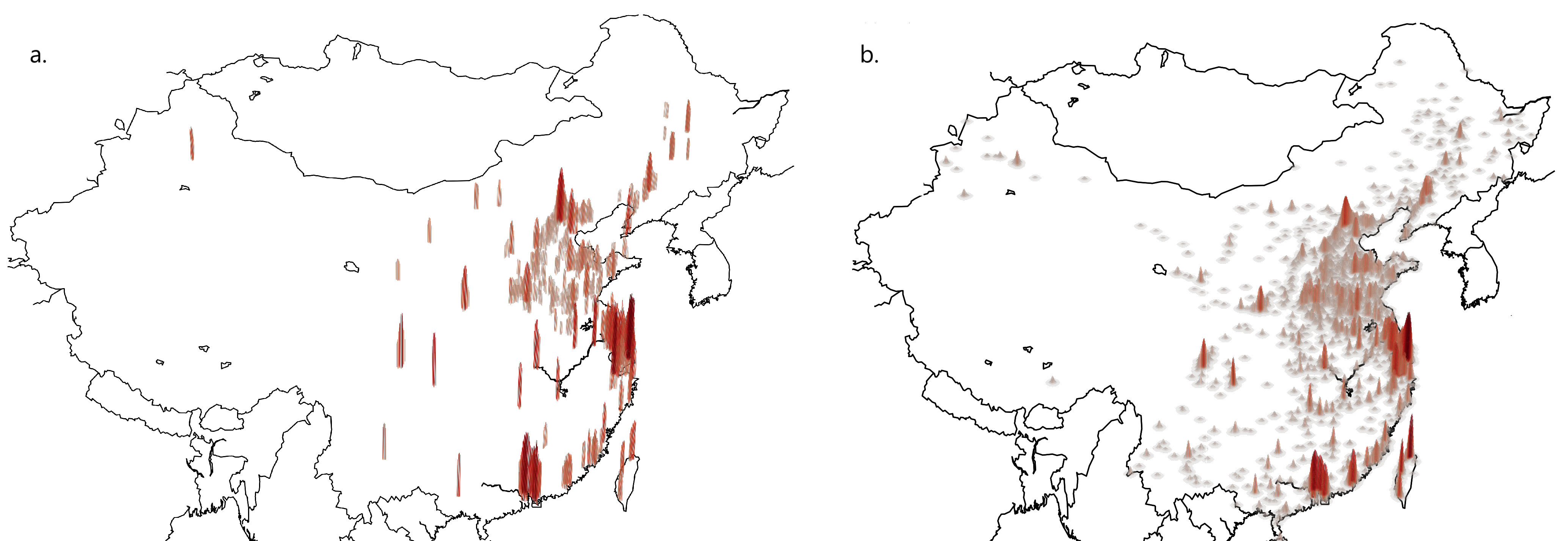}
\caption{Large-scale simulation example. (a) Building-height data from \cite{frolking2024global} for the year 2020 at a \(\sim 5\) km grid resolution, shown for cities larger than \(250\,\text{km}^2\). (b) Large-scale simulation based on Eq.~\eqref{eq:GBM_interactions} on a \(500\) m grid, initialized in 1993 and run for all cities (not only those \(>250\,\text{km}^2\)). Results are shown for the parameter set \((\mu,\sigma,k)=(0.06,0.05,0.01)\).}

    \label{fig_si:china_simu_high_res}
\end{figure}

\begin{figure}
    \centering
    \includegraphics[width=1\linewidth]{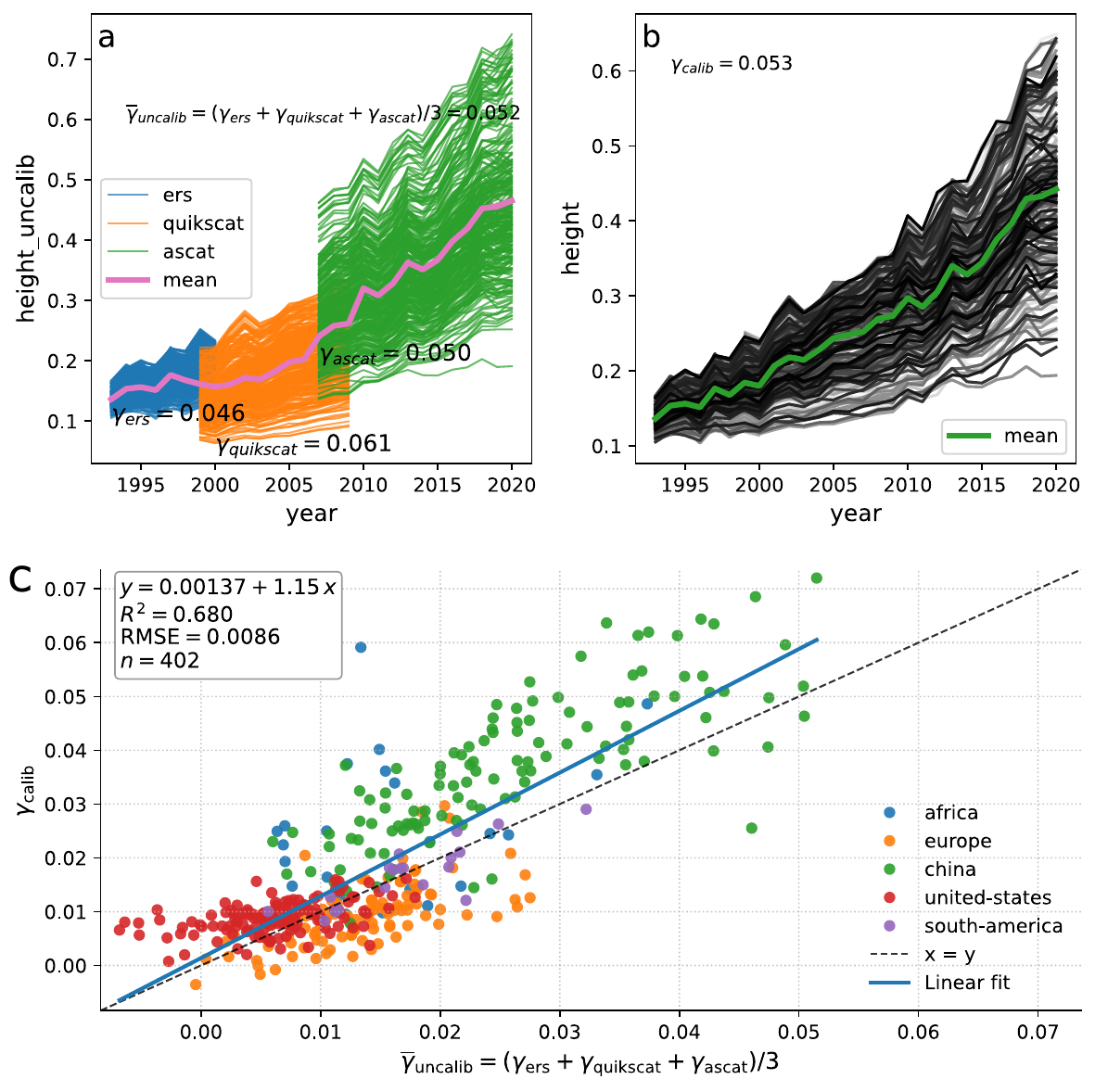}
\caption{Guangzhou intercalibration illustation. \textbf{(a)} Backscatter power-return ratio versus time before intercalibration. For each satellite segment we report the within-window $\gamma_{\text{calib}}$ and the average of the three segment $\overline{\gamma}_{\text{uncalib}}$ as an approximation  1993-2020. \textbf{(b)} Pixel time series after intercalibration. The harmonized series forms a continuous trajectory and supports a stable city-wide slope over 1993–2020. \textbf{(c)} Global comparison of \(\overline{\gamma}_{\text{uncalib}}\) versus \(\gamma_{\text{calib}}\) across cities, indicating that intercalibration has minimal impact on \(\gamma\) while aligning satellite levels.}    \label{fig_si:guangzhou_calib}
\end{figure}

\begin{figure}
    \centering
    \includegraphics[width=1\linewidth]{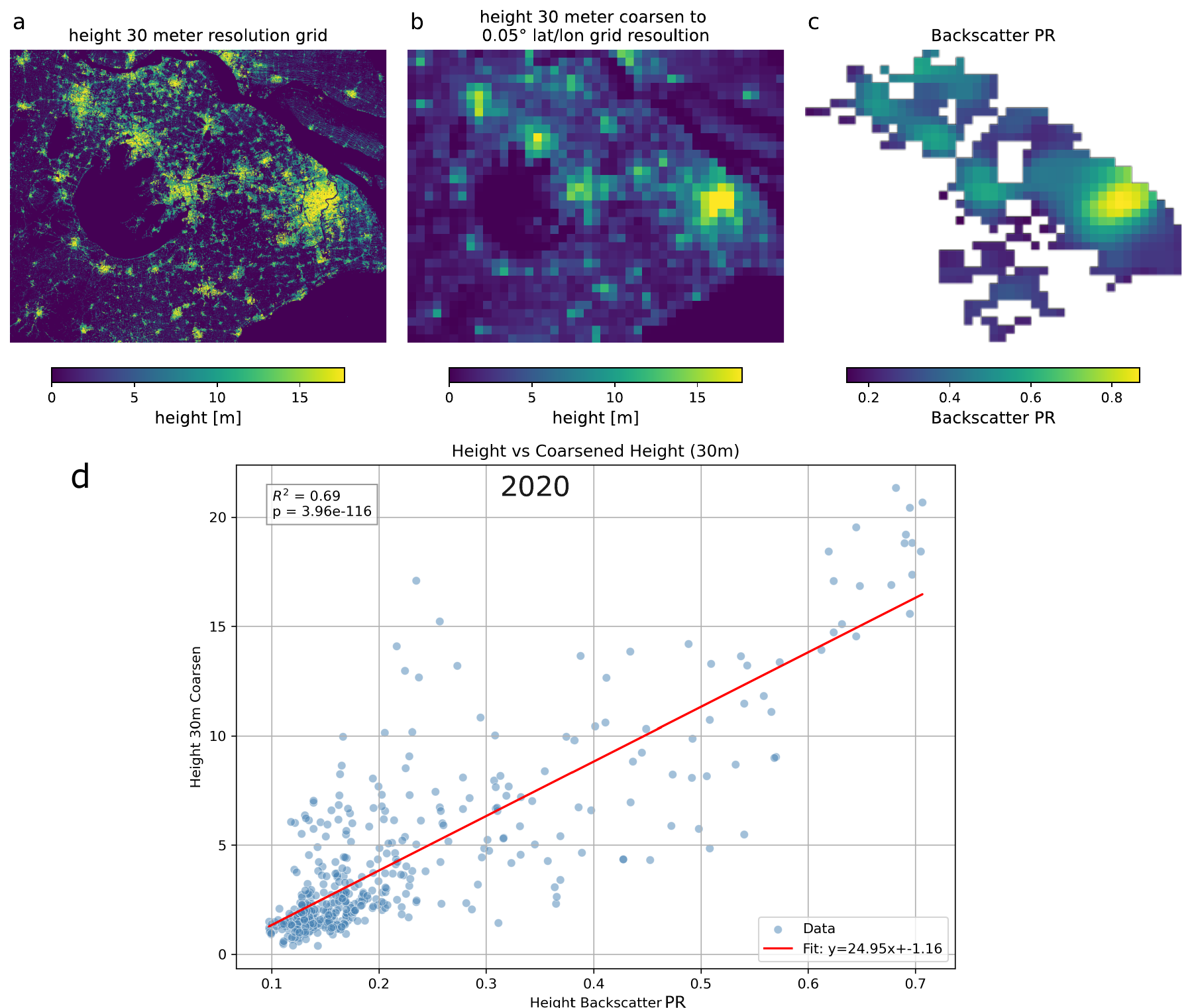}
\caption{Coarsening and cross-dataset comparison (Beijing) for year 2020. \textbf{(a)} Building height at 30\,m resolution. \textbf{(b)} The same product aggregated to the \(0.05^\circ\) grid by averaging 30\,m pixels within each cell (benchmark year 2020). \textbf{(c)} Radar backscatter power-return ratio (PR) at the \(0.05^\circ\) grid. The coarsening preserves large-scale spatial structure, and the radar field at analysis resolution shows plausible correspondence with aggregated building height. \textbf{(d)} PR--height in meters calibration for Beijing. Points are $0.05^{\circ}$ grid cells within the MUA; PR is regressed against coarsened CMTBH-30 height.}
    \label{fig_si:30m_vs_frokling}
\end{figure}

\begin{figure}
    \centering
    \includegraphics[width=1\linewidth]{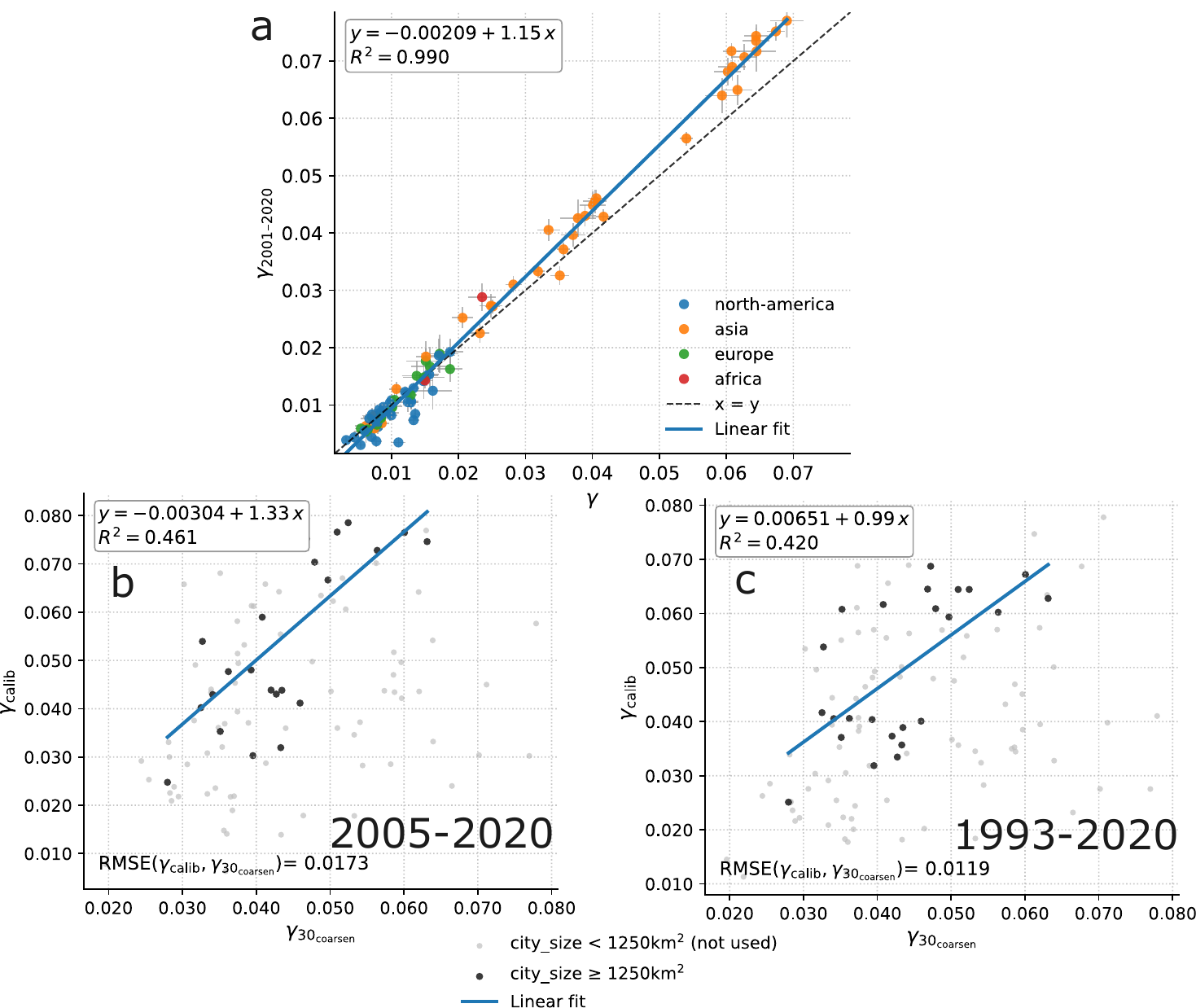}
\caption{Validation of city growth rates \(\gamma\). \textbf{(a)} Comparison of \(\gamma\) from 1993–2020 (after intercalibration) versus 2001–2020 (after intercalibration), demonstrating robustness to excluding the ERS segment with short overlap. \textbf{(b)} Comparison of radar-based \(\gamma\) from 2005–2020 (after intercalibration) with \(\gamma\) computed from the 30\,m dataset \cite{chen2025characterizing} coarsened to \(0.05^\circ\) for Chinese cities; the ranges agree well for cities larger than 1250\,km\(^2\) (approximately more than 50 grid cells). \textbf{(c)} Same as panel (b) but using radar \(\gamma\) from 1993–2020, yielding consistent conclusions.}    \label{fig_si:gamma_validation}
\end{figure}


\begin{figure}[h]
    \centering
    \includegraphics[width=0.8\linewidth]{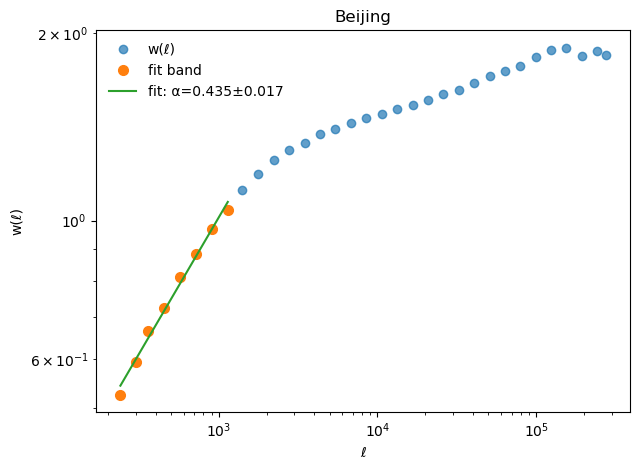}
    \caption{Illustration of the pre-plateau computation of the scaling exponent $\alpha$.}
    \label{fig:pre_plateau}
\end{figure}

\clearpage
\subsection*{Stationary PDF of the spatially coupled GBM in the mean-field approximation}

\paragraph{Stochastic mean-field model.}
We consider
\begin{equation}
\begin{aligned}
\mathrm{d}h(t) \quad &= \big[(\mu-k)h(t) + k\,m(t)\big]\,\mathrm{d}t
   + \sigma\,h(t)\,\mathrm{d}W_t, \\
m(t) \quad &:= \mathbb{E}[h(t)], \quad h(0) = h_0.
\end{aligned}
\label{eq_si:sde-h}
\end{equation}
with deterministic drift $\mu$, multiplicative noise amplitude $\sigma>0$, and constant $k>0$. Taking expectations in \eqref{eq_si:sde-h} yields
\begin{equation}
\dot m(t)=\mu\,m(t),\qquad m(t)=m_0\,\mathrm e^{\mu t},\quad m_0:=m(0).
\label{eq_si:mean}
\end{equation}

\paragraph{Centering transform and reduced SDE.}
Define the normalized process $u(t):=h(t)/m(t)$. Itô’s lemma gives
\begin{equation}
\mathrm{d}u(t)=-k\big(u(t)-1\big)\,\mathrm{d}t+\sigma\,u(t)\,\mathrm{d}W_t,\qquad
u(0)=u_0:=h_0/m_0,
\label{eq_si:sde-u}
\end{equation}
so that $h(t)=m(t)\,u(t)$ and, for any $x>0$,
\begin{equation}
p_{h(t)}(x)=\frac{1}{m(t)}\,p_{u(t)}\left(\frac{x}{m(t)}\right).
\label{eq_si:rescale}
\end{equation}
A strong solution of \eqref{eq_si:sde-u} is
\begin{equation}
u(t)=\exp\big[(-k-\tfrac12\sigma^2)t+\sigma W_t\big]\left(u_0+k\int_0^t \exp\big[(k+\tfrac12\sigma^2)s-\sigma W_s\big]\,\mathrm{d}s\right).
\label{eq_si:explicit-u}
\end{equation}
Thus the finite-time law of $u(t)$ is an affine functional of an exponential functional of Brownian motion, for which no elementary closed-form PDF is available in general. Below we provide exact moments and the stationary law.

\paragraph{Fokker--Planck equations and zero-flux stationarity.}
Let $p_h(x,t)$ denote the density of $h(t)$. The Fokker--Planck equation associated with \eqref{eq_si:sde-h} is
\begin{equation}
\begin{aligned}
\partial_t p_h(x,t) \quad
&= -\partial_x\Big( [(\mu-k)x + k\,m(t)]\,p_h(x,t) \Big) \\
&\quad + \frac{1}{2}\,\partial_{xx}\Big( \sigma^2 x^2\,p_h(x,t) \Big), \\
\text{with} \quad m(t) 
&= \int_0^\infty x\,p_h(x,t)\,\mathrm{d}x.
\end{aligned}
\label{eq_si:fp-h}
\end{equation}
For the normalized variable $u(t)$, the density $p(u,t)$ satisfies
\begin{equation}
\partial_t p(u,t)
= -\partial_u\big(-k(u-1)\,p(u,t)\big)
+\frac{1}{2}\,\partial_{uu}\big(\sigma^2 u^2\,p(u,t)\big).
\label{eq_si:fp-u}
\end{equation}
Writing \eqref{eq_si:fp-u} as a continuity equation $\partial_t p+\partial_u J=0$ defines the probability flux
\begin{equation}
\begin{aligned}
J(u,t) \quad &= a(u)\,p(u,t)
 - \tfrac{1}{2}\,\partial_u\big(b^2(u)\,p(u,t)\big), \\
a(u) \quad &= -k(u-1), \quad b(u) = \sigma u.
\end{aligned}
\label{eq_si:flux}
\end{equation}

\paragraph{Derivation of the zero-flux stationary density.}
Setting $J(u)=0$ in \eqref{eq_si:flux} yields
\begin{equation}
\begin{aligned}
\partial_u\!\big(b^2(u)\,p(u)\big) 
&= 2a(u)\,p(u) \\
\Rightarrow
\partial_u\ln\!\big(b^2(u)\,p(u)\big) 
&= \frac{2a(u)}{b^2(u)} \\
&= -\frac{2k}{\sigma^2}\left(\frac{1}{u} - \frac{1}{u^2}\right).
\end{aligned}
\end{equation}
Integrating gives
\begin{equation}
\begin{aligned}
b^2(u)\,p(u)&=C\,u^{-2k/\sigma^2}\,\exp\!\left(-\frac{2k}{\sigma^2}\,\frac{1}{u}\right)
\\ \Rightarrow\quad
p_\infty(u)&=\frac{C}{\sigma^2}\,u^{-2-2k/\sigma^2}\,
\exp\!\left(-\frac{2k}{\sigma^2}\,\frac{1}{u}\right).
\end{aligned}
\end{equation}
With $r:=\sigma^2/(2k)$ and normalization, this becomes
\begin{equation}
p_{\infty}(u)=\frac{(1/r)^{\,1+1/r}}{\Gamma(1+1/r)}\,u^{-(2+1/r)}
\exp\!\left(-\frac{1}{r\,u}\right),\qquad u>0,
\label{eq_si:stat-u}
\end{equation}
i.e.\ an inverse-gamma distribution with shape $1+1/r$ and scale $1/r$.

\paragraph{Moments and regime parameter.}
Let $\mathbb E[u_0]=1$ (i.e.\ $\mathbb E[h_0]=m_0$). From Itô on $u^2$,
\begin{equation}
\begin{aligned}
\frac{\mathrm d}{\mathrm dt}\,\mathbb E[u^2(t)]&=(-2k+\sigma^2)\,\mathbb E[u^2(t)]+2k,
\\
\Rightarrow\quad
\mathbb E[u^2(t)]&=\mathrm e^{-2k(1-r)t}\left(\mathbb E[u_0^2]-\frac{1}{1-r}\right)+\frac{1}{1-r},
\label{eq_si:m2}
\end{aligned}
\end{equation}
for $r\neq1$ (and $\mathbb E[u^2(t)]=\mathbb E[u_0^2]+2kt$ when $r=1$). Hence, with $\mathbb E[u(t)]=1$,
\begin{equation}
\mathrm{Var}[u(t)]=
\begin{cases}
\dfrac{r}{1-r}\Big(1-\mathrm e^{-2k(1-r)t}\Big), & r\neq1, \mathbb E[u_0^2]=1,\\
2kt, & r=1,\ \mathbb E[u_0^2]=1.
\end{cases}
\label{eq:var-u}
\end{equation}
\paragraph{Regimes in terms of $r=\sigma^2/(2k)$.}
\begin{itemize}
\item Subcritical noise ($r<1$). The second moment converges:
$\mathrm{Var}[u(t)]\to r/(1-r)$ as $t\to\infty$; the normalized process $u(t)$ converges in distribution to \eqref{eq_si:stat-u}, which has finite variance. Consequently, $h(t)=m(t)\,u(t)$ has a ``shape-stable'' distribution whose scale grows like $m(t)=m_0\,\mathrm e^{\mu t}$.

\item Critical noise ($r=1$). The variance grows linearly, $\mathrm{Var}[u(t)]=2kt$, and the invariant density \eqref{eq_si:stat-u} exists but has infinite variance (shape parameter $=2$).

\item Supercritical noise ($r>1$). The second moment explodes exponentially,
\begin{equation}
\mathrm{Var}[u(t)]=\frac{r}{r-1}\Big(\mathrm e^{\,2k(r-1)t}-1\Big),
\end{equation}
and the invariant density \eqref{eq_si:stat-u} has infinite variance (but remains normalizable). In this regime, tails are heavy enough that finite-time moments of order $\ge2$ eventually become dominated by rare events.
\end{itemize}

\subsection*{Inter-satellite bias correction}

We convert C-/Ku-band backscatter $\sigma^{\circ}$ (dB) to the unitless power-return ratio $h$ on the native $0.05^{\circ}$ grid, as reported in \cite{frolking2024global}, using only May composites to minimize seasonality:
\[
h^{\text{May}}_{c,i,s,t} \;=\; 10^{\,\sigma^{\circ}_{c,i,s,t,\text{May}}/10},
\]
where $c$ indexes cities, $i$ grid cells, $s$ satellites, and $t$ calendar years (1993--2020). Time is referenced to the city-specific first observation, $t \leftarrow \text{year}-t^{(c)}_0$, with $t^{(c)}_0=\min\{\text{year in city }c\}$. We analyze log-intensity $x^{(c)}_{i,s,t}=\log h^{(c)}_{i,s,t}$. The three satellites cover partially overlapping periods:
\[
\text{ERS: }1993\text{--}2000,\qquad
\text{QSCAT: }1999\text{--}2009,\qquad
\text{ASCAT: }2007\text{--}2021,
\]
yielding overlap windows $\text{ERS}\leftrightarrow\text{QSCAT}:\ 1999\text{--}2000$ and $\text{QSCAT}\leftrightarrow\text{ASCAT}:\ 2007\text{--}2009$. We take QSCAT as the reference, denoted $s^\star=\text{QSCAT}$.

\paragraph{Per-pixel overlap intercalibration (ASCAT$\leftrightarrow$QSCAT).}
For each city $c$ and pixel $i$ observed by ASCAT and QSCAT in the overlap years $\mathcal T_{i,\text{ASCAT}\leftrightarrow s^\star}=\{2007,2008,2009\}$, we use a pixel-specific affine relation on the log scale
\[
x^{(c)}_{i,\text{ASCAT},t} \;=\; a^{(c)}_{i,\text{ASCAT}} \;+\; b^{(c)}_{i,\text{ASCAT}} \, x^{(c)}_{i,s^\star,t} \;+\; u^{(c)}_{i,t},
\qquad t\in\mathcal T_{i,\text{ASCAT}\leftrightarrow s^\star}.
\]
Here, $a^{(c)}_{i,\text{ASCAT}}$ (intercept) and $b^{(c)}_{i,\text{ASCAT}}$ (slope) are unknown pixel-specific parameters, and $u^{(c)}_{i,t}$ is the residual error. We estimate $a^{(c)}_{i,\text{ASCAT}}$ and $b^{(c)}_{i,\text{ASCAT}}$ by ordinary least squares on the overlap years, and denote the estimates by $\widehat a^{(c)}_{i,\text{ASCAT}}$ and $\widehat b^{(c)}_{i,\text{ASCAT}}$.

\paragraph{Per-pixel intercept alignment (ERS$\leftrightarrow$QSCAT).}
Because the ERS–QSCAT overlap (1999–2000) is short, we align ERS to QSCAT at the pixel level using an intercept-only shift:
\[
x^{(c)}_{i,\text{ERS},t} \;=\; a^{(c)}_{i,\text{ERS}} \;+\; x^{(c)}_{i,s^\star,t} \;+\; v^{(c)}_{i,t},
\qquad t\in\{1999,2000\}.
\]
Here, \(a^{(c)}_{i,\text{ERS}}\) is an unknown pixel-specific intercept (level offset) and \(v^{(c)}_{i,t}\) is the residual error. We estimate \(a^{(c)}_{i,\text{ERS}}\) on the available overlap year(s); with one year it reduces to the log-difference \(x^{(c)}_{i,\text{ERS},t}-x^{(c)}_{i,s^\star,t}\), and with two years it is the average of those differences. The estimate is denoted by \(\widehat a^{(c)}_{i,\text{ERS}}\).

\paragraph{Applying per-pixel calibration.}
The estimated per-pixel mappings are applied to all years to place each satellite on the QSCAT scale:
\begin{equation}
\label{eq:px-apply}
\log \tilde h^{(c)}_{i,s,t} \;=\;
\begin{cases}
x^{(c)}_{i,s^\star,t}, & s=s^\star \;\;(\text{QSCAT}),\\[4pt]
\big(x^{(c)}_{i,\text{ASCAT},t} - \widehat a^{(c)}_{i,\text{ASCAT}}\big)\big/\widehat b^{(c)}_{i,\text{ASCAT}}, & s=\text{ASCAT},\\[6pt]
x^{(c)}_{i,\text{ERS},t} - \widehat a^{(c)}_{i,\text{ERS}}, & s=\text{ERS}.
\end{cases}
\end{equation}

\paragraph{Validation of the intercalibration}

We provide several figures to illustrate and validate the intercalibration of the multi-satellite backscatter record from \cite{frolking2024global} (that we refers as the radar data).

Figure~\ref{fig_si:guangzhou_calib} contrasts raw and calibrated time series for Guangzhou. Before calibration, satellite segments occupy distinct level regimes with short overlap windows. For each satellite \(s\), panel~(a) reports the within-window \(\gamma_{s}\) and the average over segments \(\overline{\gamma}_{\text{uncalib}}=(\gamma_{\text{ers}}+\gamma_{\text{quikscat}}+\gamma_{\text{ascat}})/3\) as a rough 1993–2020 approximation; this closely matches the calibrated city-wide value \(\gamma_{\text{calib}}\) (for 1993-2020 window) obtained after intercalibration in panel~(b). Panel~(c) compares \(\overline{\gamma}_{\text{uncalib}}\) to \(\gamma_{\text{calib}}\) across all cities, showing that intercalibration has little effect on \(\gamma\) and mainly harmonizes levels across satellites while preserving growth rates.

Figure~\ref{fig_si:30m_vs_frokling} illustrate the coarsening and dataset comparability with \cite{chen2025characterizing} that we use for comparison of the radar data. We plotted the 30\,m building height for Beijing, the same product aggregated to our analysis grid, and the radar power-return at that grid. The coarsening step preserves the large-scale spatial patterns, and the radar field at the coarser resolution exhibits correspondence with the aggregated building heights.

Figure~\ref{fig_si:gamma_validation} evaluates \(\gamma\) robustness. Panel~(a) compares \(\gamma\) estimated on 1993–2020 against \(\gamma\) estimated on 2001–2020 (both after intercalibration). The close agreement indicates that excluding the ERS segment—which has a very short overlap with QuikSCAT—does not alter growth-rate inference. Panel~(b) compares \(\gamma\) from 2005–2020 (radar, calibrated) against \(\gamma\) computed from the 30\,m dataset \cite{chen2025characterizing} coarsened to \(0.05^\circ\) (see Figure~\ref{fig_si:30m_vs_frokling}). Despite differing data sources and temporal sampling (2005, 2010, 2015, 2020 for the 30\,m product), the city-level growth ranges are similar, with good concordance for larger cities (area \(>\) 1250\,km\(^2\), approximately \(>\) 50 grid cells). Panel~(c) shows the same comparison using radar \(\gamma\) from 1993–2020, yielding consistent conclusions. 

Together, these indicate that the intercalibration produces stable, cross-satellite city trajectories and that the inferred growth rates are consistent with an independent, high-resolution benchmark.

\subsection*{Distance to a Theoretical Stationary Law via Jensen--Shannon Divergence}

Let $h_{c,y,i}$ denote the height for observation $i$ in city $c$ and year $y$.
To isolate distributional shape from level changes, we normalize heights within each city--year:
$$
u_{c,y,i} \;=\; \frac{h_{c,y,i}}{\overline{h}_{c,y,\cdot}}.
$$
Under the stationary hypothesis (Eq.\ref{eq_si:stat-u}), the normalized random variable $U$ follows an inverse–gamma law that depends on a single parameter $r>0$,
\begin{equation}\label{eq:stationary_u}
p_\infty(u\mid r)
= \frac{(1/r)^{\,1+1/r}}{\Gamma\!\left(1+\tfrac{1}{r}\right)}\,
  u^{-\left(2+\tfrac{1}{r}\right)}\,
  \exp\!\left(-\frac{1}{r\,u}\right),
\qquad u>0,
\end{equation}
i.e., $U\sim\mathrm{InvGamma}(\alpha, \beta)$ with $\alpha=1+1/r$ and $\beta=1/r$.

\paragraph{Estimating the stationary reference.}
For each city $c$, we estimate $\hat r_c$ from the last available year 
$y_{\max}(c) = 2020$ by maximum likelihood, equivalently fitting
$U\sim\mathrm{InvGamma}(\alpha_c,\beta_c)$ with the reparametrization
$\alpha_c=1+1/\hat r_c$ and $\beta_c=1/\hat r_c$.
This yields a time–invariant theoretical stationary density $p_\infty(\cdot\mid \hat r_c)$ used as reference for that city.

\paragraph{Empirical year-specific distributions.}
For each city--year $(c,y)$ we form an empirical distribution of $U$ on a shared grid of bins (per city):
the bin range is set to the pooled $1$–$99$th percentiles of $\{u_{c,y,i}\}_y$, and the number of bins is chosen by the Freedman--Diaconis rule \cite{freedman1981histogram}.
Let $\mathbf P_{c,y}=\{p_{c,y,k}\}_k$ be the empirical bin probabilities (histogram counts normalized so that $\sum_k p_{c,y,k}=1$).
Let $\mathbf Q_c=\{q_{c,k}\}_k$ be the corresponding bin probabilities obtained by integrating the fitted density,
$q_{c,k}=\int_{\mathrm{bin}\ k} p_\infty(u\mid \hat r_c)\,\mathrm du$ (computed from the inverse–gamma CDF).

\paragraph{Jensen--Shannon divergence (JSD).}
We quantify the dissimilarity between the empirical year-$y$ distribution and the stationary reference using the base-2 Jensen--Shannon divergence \cite{lin2002divergence}:
$$
\mathrm{JSD}_2(\mathbf P_{c,y}\,\Vert\,\mathbf Q_c)
\;=\;
\tfrac{1}{2}\,\mathrm{KL}_2\!\big(\mathbf P_{c,y}\,\Vert\,\mathbf M_{c,y}\big)
\;+\;
\tfrac{1}{2}\,\mathrm{KL}_2\!\big(\mathbf Q_c\,\Vert\,\mathbf M_{c,y}\big),
\quad
\mathbf M_{c,y}=\tfrac{1}{2}(\mathbf P_{c,y}+\mathbf Q_c),
$$
where $\mathrm{KL}_2$ uses $\log_2$.
Thus $\mathrm{JSD}_2\in[0,1]$, with $0$ indicating identical distributions.

\paragraph{Trend and short-horizon diagnostics.}
For each city we obtain a time series $d_{c,y}=\mathrm{JSD}_2(\mathbf P_{c,y}\Vert \mathbf Q_c)$ over years.
To summarize convergence toward the stationary law, we estimate a linear trend
$s_c$ (slope per year) of $d_{c,y}$ vs.\ $y$ ; a negative slope indicates decreasing divergence (convergence).

\subsection*{Local-width roughness on a log–height field with masking and pre-plateau fitting}

\paragraph{Log transform and validity mask.}
Starting from a height field $h(x,y)$, we analyze the log–height field
$$
\phi(x,y)\;=\;\log\!\big(h(x,y)+\varepsilon\big)\;-\;\min\nolimits_{(i,j)\in\mathcal{V}} \log\!\big(h_{ij}+\varepsilon\big)\;+\;\delta,
$$
with a small offset $\varepsilon>0$ (chosen a priori; $\varepsilon=0.01\,\mathrm{median}\{h>0\}$) and a tiny shift $\delta>0$ ($\delta=10^{-6}$) to make $\phi>0$ on valid pixels. Pixels where $h+\varepsilon\le 0$ are treated as holes (set to $0$ in memory but excluded by the mask). The set of valid pixels is
$$
\mathcal{V}=\{(x,y):\, h(x,y)+\varepsilon>0,\ \text{$h$ finite}\}.
$$
All statistics below are computed exclusively over $\mathcal{V}$. Because the local mean is subtracted inside each box (see below), the global constant shift $-\min\log(h+\varepsilon)+\delta$ does not affect the local-width values.

\paragraph{Local-width.}
For a square box of side $\ell$ centered at $(x,y)$, let $B_\ell(x,y)\subset\mathcal{V}$ be the set of valid pixels in the box. The local width on $\phi$ is the RMS after subtracting the local mean:
\begin{align*}
w_\ell^2(x,y) &= \frac{1}{|B_\ell(x,y)|}\sum_{(i,j)\in B_\ell(x,y)}\Big(\phi_{ij}-\bar \phi_\ell(x,y)\Big)^2,\\
\bar \phi_\ell(x,y) &= \frac{1}{|B_\ell(x,y)|}\sum_{(i,j)\in B_\ell(x,y)} \phi_{ij}.
\end{align*}
Boxes are accepted only if their coverage exceeds a threshold,
$$
\frac{|B_\ell(x,y)|}{\ell^2}\ \ge\ \texttt{min\_coverage}\ \ (\text{set to\ }0.8),
$$
otherwise they are discarded. The scale-aggregated width is the across-box average
$$
w(\ell)=\big\langle w_\ell(x,y)\big\rangle_{\text{boxes at size }\ell}.
$$
We estimate $w(\ell)$ on a geometric ladder $\ell\in\{\ell_0,\ \ell_0 r,\ \ell_0 r^2,\dots\}$ with ratio $r\approx1.25$, from \texttt{min\_box} up to \texttt{max\_box} (default $\simeq L/4$, where $L$ is the smaller image side). For each $\ell$ we also record the sample standard deviation across boxes, $s_w(\ell)$, and the number of contributing boxes, $n(\ell)$.

\paragraph{Pre-plateau extraction of the roughness exponent $\alpha$ (see Fig.\ref{fig:pre_plateau} for an illustration).}
For a self-affine field, the local width scales as
$$
w(\ell)\ \sim\ \ell^{\alpha}\qquad (a\ll \ell \ll L),
$$
where $a$ is the pixel size and $L$ the system size. At large $\ell$ finite-size effects induce a plateau. We therefore fit only the pre-plateau band via:
\begin{enumerate}
\item \textbf{Log--log variables:} $x_i=\log \ell_i$, $y_i=\log w(\ell_i)$ for scales with $w(\ell_i)>0$ and $n(\ell_i)>0$.
\item \textbf{Plateau detection by local slopes:} compute a sliding OLS slope $s_i$ of $y$ vs.\ $x$ over $w$ points (set $w=5$). The plateau onset index $i^\ast$ is the first position where two consecutive slopes satisfy $s_i<s_{\rm flat}$ and $s_{i+1}<s_{\rm flat}$ (default $s_{\rm flat}=0.3$). The fit band is $[i_0,i_1]$ with $i_0$ skipping the smallest scales (in this study we used \texttt{ignore\_first}$=0$ i.e. we do not skip the first box) and $i_1=\max(i^\ast-1,\ i_0+\texttt{min\_points}-1)$. If no plateau is detected, fit up to the largest scale; require at least \texttt{min\_points} (set to $6$).
\item \textbf{Weighted least squares:} on $[i_0,i_1]$, estimate
$$
y=\alpha\,x+b+\varepsilon
$$
by WLS with weights
$$
w_i\ \propto\ \frac{n(\ell_i)}{\left[s_w(\ell_i)/w(\ell_i)\right]^2},
$$
i.e., inverse-variance weighting in $\log w$ scaled by the number of boxes. Report $\hat{\alpha}$ and its standard error from the WLS covariance.
\end{enumerate}

\end{document}